\documentclass[aps,twocolumn,superscriptaddress,amsmath,amssymb,nofootinbib]{revtex4-1}

\usepackage{graphicx}
\usepackage{dcolumn}
\usepackage{bm}
\usepackage{color}
\usepackage{txfonts}
\usepackage{hyperref}
\usepackage{soul}
\usepackage{xr}
\definecolor{dkgreen}{rgb}{0,0.6,0}
\definecolor{gray}{rgb}{0.5,0.5,0.5}
\definecolor{mauve}{rgb}{0.58,0,0.82}

\newcommand{\rT}{\textrm{\tiny T}}

\begin{document}
\title{Coherence enhanced quantum-dot heat engine}

\author{Jaegon Um}
\email{slung@postech.ac.kr}
\affiliation{Department of Physics, Pohang University of Science and Technology, Pohang 37673, Korea}

\author{Konstantin E.~Dorfman}
\email{dorfmank@lps.ecnu.edu.cn}
\affiliation{State Key Laboratory of Precision Spectroscopy, East China Normal University,
Shanghai 200062, China}

\author{Hyunggyu Park}
\email{hgpark@kias.re.kr}
\affiliation{School of Physics, Korea Institute for Advanced Study, Seoul 02455, Korea}
\affiliation{Quantum Universe Center, Korea Institute for Advanced Study, Seoul 02455, Korea}


\begin{abstract}

We show that quantum coherence can enhance the performance of a continuous quantum heat engine
in the Lindblad description.
We investigate the steady-state solutions of the particle-exchanging
quantum heat engine, composed of
degenerate double quantum dots coupled to two heat baths in parallel,
where quantum coherence may be induced due to interference between relaxation channels.
We find that the engine power can be enhanced by the coherence in the nonlinear response regime,
when the symmetry of coupling configurations between dots and two baths is broken. In the symmetric
case, the coherence cannot be maintained in the steady state, except for the maximum interference degenerate case,
where initial-condition-dependent multiple steady states appear with a dark state.
\end{abstract}



\maketitle

\emph{Introduction} -- Quantum thermodynamics is an emerging field in view of significant
progress of technology which allows to scale down heat-energy
converting devices to nanoscale where quantum effects cannot be
neglected~\cite{ben17}.
Examples of such quantum heat engines (QHE) include lasers,
solar cells, photosynthetic organisms, etc where along with
few-level quantum structure~\cite{scu03,kos14,jar16}
a phenomenon of quantum coherence plays an important
role~\cite{che16,whi16,aga17}.
In particular,
coherence in system-bath interactions that originates from the interference
may enhance the power~\cite{scu11,yao15} and efficiency
at maximum power~\cite{dor18} of the laser and solar cell and is responsible
for highly efficient energy transfer in photosynthetic systems~\cite{Dorfman:2013PNAS}.
These effects have been confirmed in the experimental studies of polymer solar cells~\cite{bit14}.
The noise-induced coherence is a different kind from
internal coherence usually involved with a system Hamiltonian~\cite{uzdin}
which was recently demonstrated
in the nitrogen vacancy-based microscopic QHE in diamond~\cite{kla17}, and manifests
as an improved efficiency in spectroscopic pump-probe measurements~\cite{QuDorf}.

\begin{figure}
\includegraphics[width=\columnwidth]{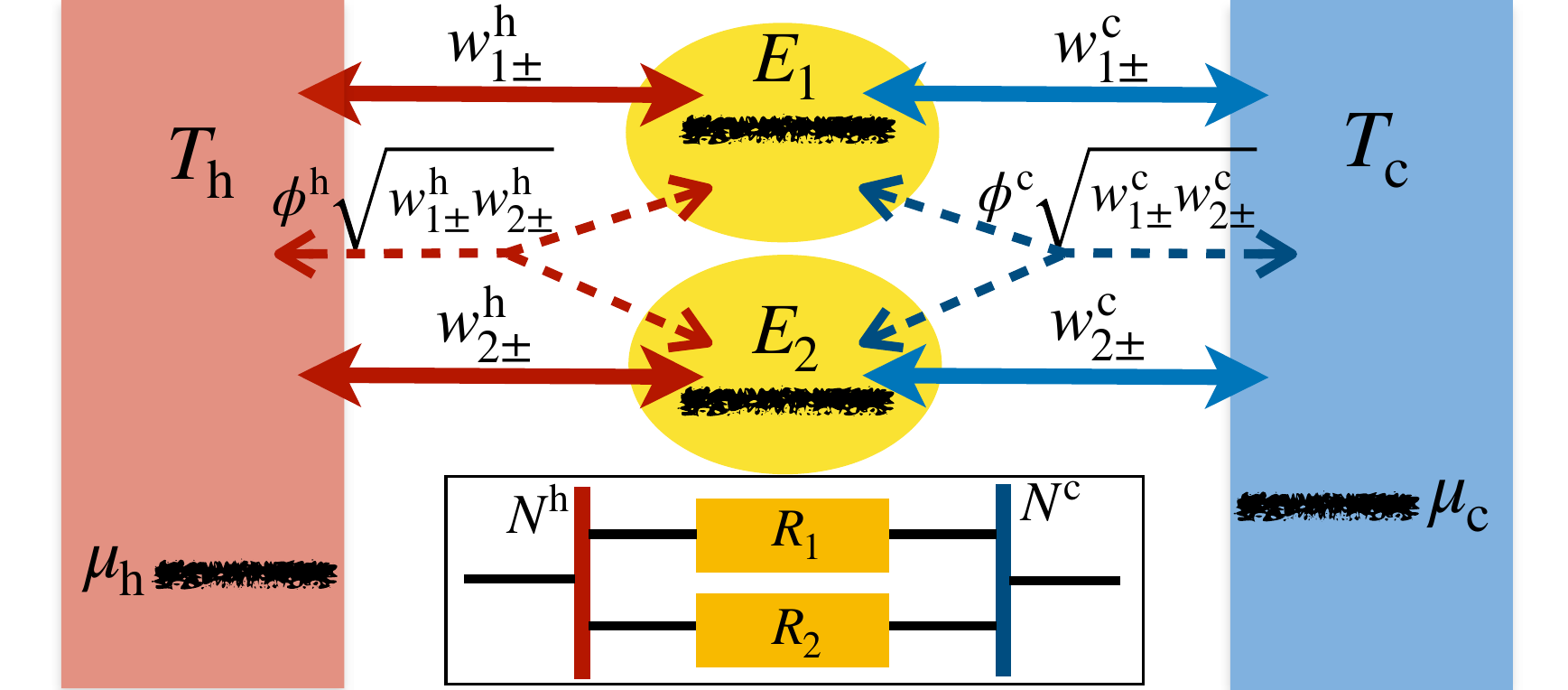}
\caption{A schematic illustration of the QHE, composed of two heat baths and a two-dot system.
The dot energies, $E_1$ and $E_2$ (in this work, $E_1=E_2$) are
higher than chemical potentials, $\mu_{\rm h}$ and $\mu_{\rm c}$.
$w^{a}_{d \pm}$ represents transfer rate of a particle between dot $d$ and bath $a$,
and $\phi^{a}
\sqrt{ w^{a}_{1 \pm} w^{a}_{2 \pm}}$ denotes the interference amplitude.
Inset: A circuit analogy of resistors in parallel.
}
\label{fig:1}
\end{figure}

In much of the literature, the quantum coherence effect has been studied on continuously-working
bosonic devices~\cite{scu11,yao15, dor18,Dorfman:2013PNAS}. In this Letter, we focus on the fermionic QHE autonomously working without external work agents like a driving laser,
made up of repulsively interacting double quantum dots with the same energy levels,
coupled to fermionic baths in parallel, depicted
in Fig.~\ref{fig:1}.
In contrast to previous studies for such
a system~\cite{harbola,schaller,cuetara}, we introduce
a parameter for the strength of interference between relaxation channels, which
plays a crucial role.
We derive the condition for maintaining quantum coherence in
the steady state and investigate the engine performance, not only in terms of tunneling
coefficients between dots and baths, but also interference strengths.

We find that the power enhancement of the QHE can be achieved in the nonlinear response regime.
When coupling configurations assigned to each bath are symmetric to each other,
a quantum coherence initially induced by interference between relaxation channels would eventually disappear in the long-time (steady-state) limit.  The exceptional case emerges for the degenerate energy level configuration at the maximum interference strength,
where the dynamics is found to be localized, manifested as a mathematical singularity in the evolution operator evoking
the so-called {\em dark} state~\cite{darkstate}, characterized by multiple steady states with finite quantum coherence depending on a given initial state.
This singularity should be also found for more general settings with coherent dynamics
originated from the energy-level degeneracy and parallel couplings, including a single bath case.
Note that a spurious quantum coherence can be observed for a very long time (quasi-stationary state regime) near the maximum interference.

When the coupling configuration symmetry is broken in terms of either tunnelling coefficients or interference strengths,
a genuine new steady state emerges with non-vanishing quantum coherence in general, producing an extra  {\em quantum current} between two baths through dots in addition to the conventional {\em classical} current. This quantum current yields an extra contribution to the engine power, which can be positive in a specific parameter regime.

\emph{Model} -- We first derive the quantum master equation (QME)~\cite{bre02}
for the density operator $\hat{\rho}_{\rm S} (t)$ of the fermionic QHE
in the limit of weak coupling to hot ($\rm h$) and cold ($\rm c$) baths,
where a temperature difference $T_{\rm h} - T_{\rm c} >0$ and
a potential bias $\mu_{\rm c} -\mu_{\rm h} >0$ are applied.
For simplicity, we assume a single energy level for each quantum dot
with the degenerate energy levels $E_1=E_2=E$ and infinitely large repulsion between
particles in dots.
The system then can be described using three two-particle eigenstates:
$|0 \rangle$ denotes empty dots, $|1 \rangle$ and $|2 \rangle$
stand for occupation of dot 1 and 2, respectively by a single particle.
In addition, coherent hopping between dots is also forbidden
and the only source of coherence is due to coupling to thermal baths.

The interaction between system and bath $a(=\rm h\,,c)$ is given by
$\hat{H}^a_{\rm SB} = \sum_{d,k} g^a_d {\hat{b}_k^{a\dagger}}
~|0 \rangle \langle d| + h.c.$, where $\hat{b}^{a\dagger}_k$ is
the operator creating a single particle with momentum $k$ in bath $a$ and
$g^a_d$ is the tunneling coefficient between dot $d(=1,2)$ and bath $a$.
Following the standard procedure of tracing out bath degrees of freedom, we obtain the QME which reads
\begin{equation}
\partial_t {\hat{\rho}}_{\rm S} = -{\rm i}[\hat{H}_{\rm S}, \hat{\rho}_{\rm S}]
+  \sum_{a}  \sum_{\alpha,\beta=1}^4\Gamma^{a}_{\alpha \beta}
\left( \hat{L}_{\alpha} \hat{\rho}_{\rm S} \hat{L}^{\dagger}_{\beta}
-\frac{1}{2}\left\{  \hat{L}^{\dagger}_{\beta} \hat{L}_{\alpha}, \hat{\rho}_{\rm S}
\right\} \right) \,.
\label{eq:lind2}
\end{equation}
where the system Hamiltonian $\hat{H}_{\rm S} = E \left( |1 \rangle \langle 1 | + |2 \rangle \langle 2 | \right)$
and the Lindblad operators are $\hat{L}_{1} =  |1 \rangle \langle 0|$, $\hat{L}_{2} =  |2 \rangle \langle 0|$,
$\hat{L}_{3} =  \hat{L}_{1}^{\dagger}$, and $\hat{L}_{4} =  \hat{L}_{2}^{\dagger}$.
Note that we neglected the Lamb shift term~\cite{supp}.
The dissipation matrix $\Gamma^a$ is
given by
\begin{equation}
\Gamma^a= \left( \begin{array}{cccc}
w^a_{1+} & \phi^a \sqrt{w^a_{1+} w^a_{2+}} & 0 & 0 \\
\phi^{a*} \sqrt{w^a_{1+} w^a_{2+} } & w^a_{2+} & 0 & 0 \\
0 & 0 & w^a_{1-} & \phi^{a*} \sqrt{w^a_{1-} w^a_{2-}}  \\
0 & 0 & \phi^{a} \sqrt{w^a_{1-} w^a_{2-}} & w^a_{2-} \end{array} \right) \,
\label{eq:gamma}
\end{equation}
where $w^a_{d\pm}$ represents transfer rate of a particle between dot $d$ and bath $a$;
the subscript $+$ ($-$) denotes the inflow (outflow) with respect to the dot, respectively.
These rates are given by
$w^a_{d+} = 2\pi \lvert g^a_{d}\rvert ^2 N^a$ and
 $w^a_{d-}=  2\pi \lvert g^a_{d}\rvert ^2 \overline{{N}^{a}} $, where
$N^a = N^a(E)$ is the Fermi-Dirac distribution in bath $a$ and
$\overline{{N}^{a}} = 1-N^a$ (see the derivation in Sec.~\ref{sec:ME} of the Supplemental Material(SM)~\cite{supp}).

The off-diagonal terms in Eq.~\eqref{eq:gamma}
represent interference between particle transfer associated with different dots.
The interference effect will be manifested
in nonzero off-diagonal terms of $\hat{\rho}_{\rm S}$, e.g.,
$\langle 1| \hat{\rho}_S |2 \rangle \neq 0$.
The coherence, however, can be destroyed by other environmental noises
not captured in the interaction Hamiltonian,
like the gate voltage noise~\cite{gatenoise}
which may dephase the system.
Instead of the secular approximation removing such coherence entirely
in the dissipation~\cite{sa}, we introduce a phenomenological parameter $\phi^a$
for the interference term in Eq.~\eqref{eq:gamma},
assigned to each bath; $|\phi^a|=1$ stands for
permitting the full interference of relaxations with bath $a$, while
$\phi^a=0$ corresponds to no quantum effect of system-bath interactions.
In earlier bosonic QHE models, this interference parameter is governed by the angle
between dipole moments corresponding to two dots which ensures that $|\phi^a|\leq 1$~\cite{scu11}.
For convenience, $\phi^a$ is treated as a real number.
Note that the second term in Eq.~\eqref{eq:lind2}
is a standard form of
quantum dynamical semigroup~\cite{bre02},
which guarantees the positive and trace preserving dynamics since
$\Gamma^a$  in Eq.~\eqref{eq:gamma}
is the positive-semidefinite matrix for $|\phi_a|\le 1$.

To solve the QME, it is convenient to map the density operator to
a vector: $\mathbf{P} = \left(\rho_{00}, \rho_{11}, \rho_{22}, \rho_{12}, \rho_{21},
\rho_{01}, \rho_{02}, \rho_{10}, \rho_{20} \right)^{\rT}$,
where $\rho_{ij} = \langle i | \hat{\rho}_{\rm S} | j \rangle$.
The last four components vanish in the long-time limit because
there is no dynamics producing the coherence between the empty and
occupied states so that only dephasing is allowed, as seen in Sec.~\ref{sec:eigen} of SM~\cite{supp}.
Thus, we write the corresponding Liouville equation as
\begin{equation}
\partial_t \mathbf{P} =\mathsf{L}\, \mathbf{P}~,
\label{eq:lind3}
\end{equation}
where $\mathsf{L}$ is a $5\times5$ matrix with the reduced vector
$\mathbf{P}=\left( \rho_{00}, \rho_{11}, \rho_{22}, \rho_{12}, \rho_{21} \right)^{\rT}$.
Introducing $W_{d} = \sum_a w^a_{d+}$,
$\overline{W}_{d} = \sum_a w^a_{d-}$,
$\Phi = \sum_a \phi^{a} \sqrt{w^a_{1 +}w^a_{2 +} } $, and
$\overline{\Phi} = \sum_a \phi^{a} \sqrt{w^a_{1 -}w^a_{2 -} } $,
the $\mathsf{L}$ matrix then reads
\begin{equation}
\label{eq:LW}
\mathsf{L}= \left(
\begin{array}{ccccc}
-W_1 - W_2  & \overline{W}_1 & \overline{W}_2 &  \overline{\Phi} & \overline{\Phi} \\
W_1 & -\overline{W}_1 & 0 & - \overline{\Phi} /2 & - \overline{\Phi}/2 \\
W_2 & 0 & -\overline{W}_2 & - \overline{\Phi}/2 & -\overline{\Phi}/2 \\
\Phi  &  -\overline{\Phi} /2
& -\overline{\Phi} /2 & -\frac{\overline{W}_1 + \overline{W}_2}{2} & 0 \\
\Phi & -\overline{\Phi} /2
& -\overline{\Phi}/2  & 0 & -\frac{\overline{W}_1 + \overline{W}_2}{2}
 \end{array} \right).
\end{equation}

\emph{Steady-state solutions} -- From the steady-state condition, $\mathsf{L} \mathbf{P}(\infty) =0$,
we find the relations as
\begin{eqnarray} \label{eq:sol_ss1}
\rho_{11}(\infty) &=& \frac{ W_1 \overline{W}_2 - \overline{\Phi}
\left( W_2 + \overline{W}_2 - W_1 \right) \rho_{12}(\infty) }
{W_1 \overline{W}_2 + \overline{W}_1 W_2 + \overline{W}_1\overline{W}_2 }\,, \nonumber\\
\rho_{22}(\infty) &=& \frac{  \overline{W}_1 W_2 - \overline{\Phi}
\left( W_1 + \overline{W}_1 - W_2 \right) \rho_{12}(\infty) }
{W_1 \overline{W}_2 + \overline{W}_1 W_2 + \overline{W}_1\overline{W}_2 } \,,
\end{eqnarray}
with the population conservation ($\rho_{00} + \rho_{11} +\rho_{22}=1$) and
\begin{equation} \label{eq:sol_ss2}
 \rho_{12}(\infty) =\rho_{21}(\infty)= \frac{2 \Phi - \left( 2 \Phi + \overline{\Phi} \right)
\left( \rho_{11}(\infty)  +\rho_{22}(\infty) \right)   }
{\overline{W}_1 +\overline{W}_2 } \,.
\end{equation}
Note that the classical solution is recovered from Eq.~\eqref{eq:sol_ss1}, when the coherence term vanishes ($\rho_{12}(\infty) =0$).
This classical \emph{incoherent condition} is determined by Eq~\eqref{eq:sol_ss2} as
\begin{equation}\label{eq:no_coh}
2\Phi \overline{W}_1 \overline{W}_2 - \overline{\Phi}
\left( W_1 \overline{W}_2 + \overline{W}_1 W_2 \right) =0\,
\end{equation}
which is obviously satisfied for the trivial case with $\Phi=\overline{\Phi}=0$ (or equivalently $\phi^a=0$).
Note that the equilibrium case ($T_{\rm h}=T_{\rm c}$ and $\mu_{\rm h}=\mu_{\rm c}$) also satisfies this incoherent condition
due to $W_d/\overline{W}_d=\Phi/\overline{\Phi}$ with $N^{\rm h}=N^{\rm c}$.

In general, Eqs.~\eqref{eq:sol_ss1} and \eqref{eq:sol_ss2} leads to a $2\times2$ matrix equation for $\rho_{11}$ and $\rho_{22}$ as
\begin{equation} \label{eq:ss_general}
\mathsf{L}_{\rm ss} \left( \begin{array}{c}
\rho_{11}(\infty) \\ \rho_{22}(\infty) \end{array} \right) = \left(
\begin{array}{c} W_1  - \left(2\Phi \overline{\Phi}\right)/
\left( \overline{W}_1 +\overline{W}_2 \right)   \\
W_2 - \left(2\Phi \overline{\Phi} \right)/
\left(\overline{W}_1 +\overline{W}_2  \right)  \end{array} \right) \,,
\end{equation}
with
\begin{equation} \label{eq:ss_matrix}
\mathsf{L}_{\rm ss} = \left( \begin{array}{cc}
W_1 + \overline{W}_1 -\frac{ \overline{\Phi} \left( 2 \Phi + \overline{\Phi} \right) }
{\overline{W}_1 +\overline{W}_2 }
& W_1 - \frac{ \overline{\Phi} \left( 2 \Phi + \overline{\Phi} \right) }
{\overline{W}_1 +\overline{W}_2 }   \\
W_2 - \frac{ \overline{\Phi} \left( 2 \Phi + \overline{\Phi} \right) }
{\overline{W}_1 +\overline{W}_2 }
& W_2 + \overline{W}_2 -\frac{ \overline{\Phi} \left( 2 \Phi + \overline{\Phi} \right) }
{\overline{W}_1 +\overline{W}_2 }
\end{array} \right) \,.
\end{equation}
Unless the determinant $|\mathsf{L}_{\rm ss}|$ vanishes, the steady state solution is
uniquely defined, which is given explicitly in Eq.~\eqref{eq:sol_ss3} of SM~\cite{supp}.

We next consider a special case, where
the coupling strength ratio of two dots with a bath is
the same for both baths, i.e.~$g^{\rm h}_{2}/g^{\rm h}_{1} = g^{\rm c}_{2}/g^{\rm c}_{1} \equiv r$,
leading to $w^{a}_{2\pm}/w^{a}_{1\pm} = r^2$. This may be a natural situation in real experiments
and will be called the \emph{$r$-symmetric} configuration~\cite{schaller}. We take $r>0$ for simplicity.
Assuming an additional symmetry for the coherence parameter as $\phi^{\rm h}=\phi^{\rm c}\equiv \phi$,
one can show $W_d/\overline{W}_d=\Phi/\overline{\Phi}$ even in nonequilibrium ($N^{\rm h}\neq N^{\rm c}$),
satisfying the incoherence condition in Eq.~\eqref{eq:no_coh}. However, at the maximum interference
($|\phi|=1$), the matrix $\mathsf{L}_{\rm ss}$ becomes singular with $|\mathsf{L}_{\rm ss}|=0$ and
multiple steady-state solutions emerge, which will be discussed later.
With the broken symmetry $(\phi^{\rm h}\neq \phi^{\rm c})$, the quantum coherence survives
with a non-classical solution ($\rho_{12}(\infty)\neq 0$). In a more general case with
$g^{\rm h}_{2}/g^{\rm h}_{1} \neq g^{\rm c}_{2}/g^{\rm c}_{1}$, the classical solution is
still possible by adjusting $\phi^{\rm h}$ and $\phi^{\rm c}$ appropriately to satisfy the incoherent
condition, but $\mathsf{L}_{\rm ss}$ cannot be singular.

\emph{Steady-state currents} --
A particle current $J_d^a$ representing a particle net flow to dot $d$ from bath $a$
can be obtained from the QME in Eq.~\eqref{eq:lind2} or Eq.~\eqref{eq:lind3} as
\begin{align}\label{eq:Jda}
J_d^a= w_{d+}^a ~\rho_{00}- w_{d-}^a ~\rho_{dd}-\phi^a \sqrt{w^a_{1-} w^a_{2-}} \left(\frac{\rho_{12}+\rho_{21}}{2}\right)~,
\end{align}
which represents time increment of the particle density of dot $d$  due to bath $a$.
In the steady state, we expect that the particle density increment should be balanced by two reservoirs such that
$J_d^{\rm h}(\infty)=-J_d^{\rm c}(\infty)\equiv J_d(\infty)$ and the total current is
given by $J=\sum_d J_d(\infty)$.
Transferring an electron from bath h to bath c, the electron gains
an energy governed by the difference between underlying chemical potentials $\mu_{\rm c}-\mu_{\rm h}$, thus the output power of the QHE
is simply given by $P=(\mu_{\rm c}-\mu_{\rm h}) J$.
As the heat flux from the high-temperature reservoir
is given as $\dot{Q}^{\rm h}=(E-\mu_{\rm h})J$, the engine efficiency does not vary with the particle current as
$\eta=P/\dot{Q}^{\rm h}=(\mu_{\rm c}-\mu_{\rm h})/(E-\mu_{\rm h})$.

The particle current can be further separated into the classical and quantum part as
\begin{align}\label{eq:Jd}
J_d(\infty)=J_d^{\rm cl} + \Psi_d ~\rho_{12}(\infty)~,
\end{align}
where the classical current is defined by setting $\phi^a=0$ in Eqs.~\eqref{eq:sol_ss1} and \eqref{eq:Jda} as
\begin{align}\label{eq:Jdcl}
J_1^{\rm cl}=\frac{\Delta N}{|\mathsf{L}_0|}(2\pi)^2 |g_1^{\rm h}|^2|g_1^{\rm c}|^2\overline{W}_2 \, ,
\, J_2^{\rm cl}=\frac{\Delta N}{|\mathsf{L}_0|}(2\pi)^2 |g_2^{\rm h}|^2|g_2^{\rm c}|^2\overline{W}_1
\end{align}
with the external (bath) bias $\Delta N\equiv N^{\rm h}-N^{\rm c}$ and $|\mathsf{L}_0|=
W_1 \overline{W}_2 + \overline{W}_1 W_2 + \overline{W}_1\overline{W}_2$ ($\mathsf{L}_0=\mathsf{L}_{\rm ss}(\phi^a=0)$).
For a proper classical engine to generate positive power ($J_d^{\rm cl}>0$), we consider $\Delta N>0$ only.

The second term represents the \emph{quantum current} $J_d^{\rm q}\equiv \Psi_d ~\rho_{12}(\infty)$,
induced by the coherence, and
the explicit expressions for the \emph{quantum speed} $\Psi_d$ and $\rho_{12}(\infty)$ are given in Sec.~\ref{sec:current} of SM~\cite{supp}.
Note that the quantum current for each dot can be both positive and negative, depending on the parameter values, as well
as the total quantum current $J^{\rm q}=\sum_d J_d^{\rm q}$ (see Fig.~\ref{fig:s1} of of SM~\cite{supp}).

\begin{figure}
\includegraphics[width=1.0\columnwidth]{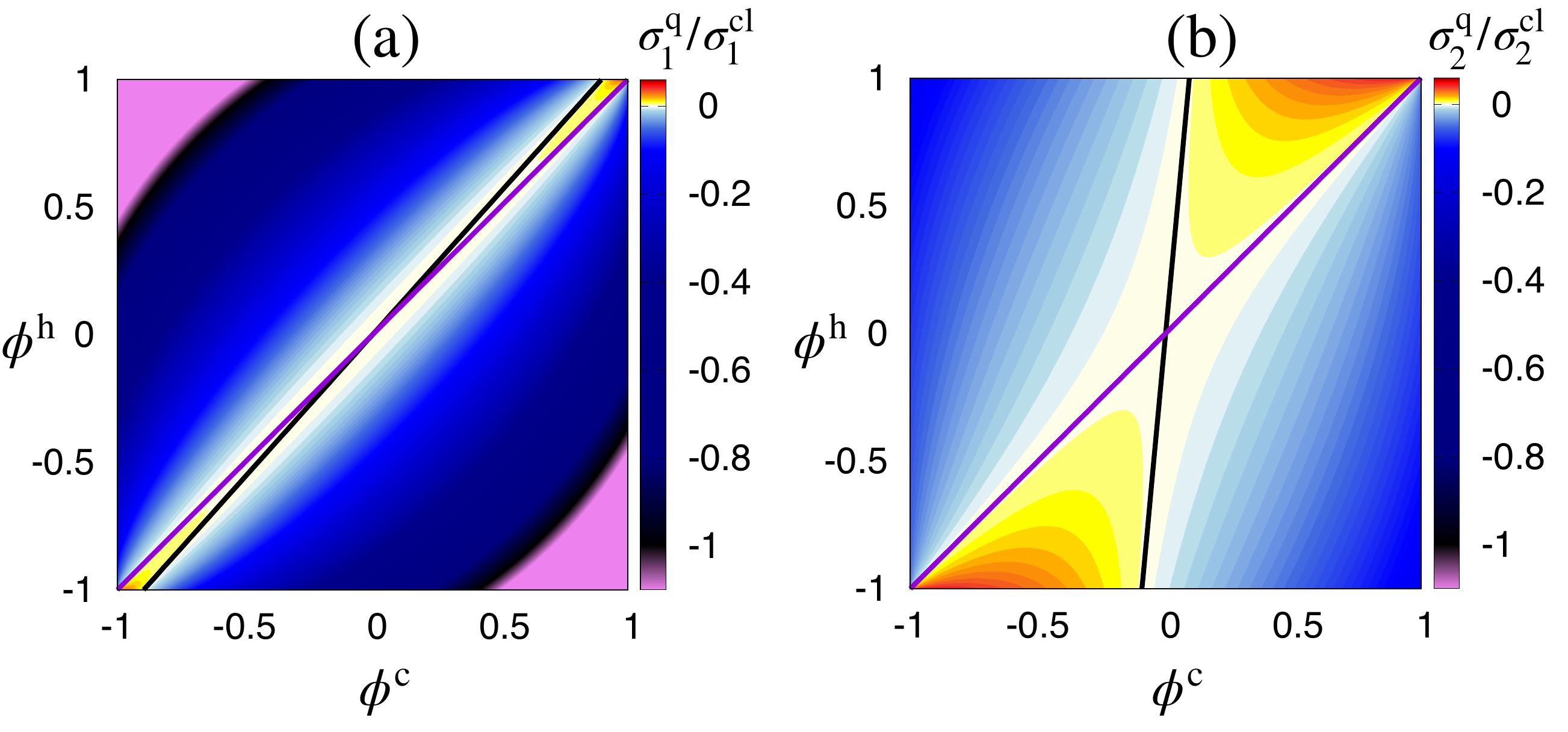}
\caption{
Relative quantum conductances of (a) dot 1 and (b) dot 2,
denoted as $\sigma^{\rm q}_1 /\sigma^{\rm cl}_1$ and
$\sigma^{\rm q}_2 /\sigma^{\rm cl}_2$, respectively in
the ($\phi^{\rm c}, \phi^{\rm h}$) plane.
Here, we used $N^{\rm h}=0.2$ and  $N^{\rm c} =0.1$,
and the $r$-symmetric configuration with $|g^a_1|^2 = 8\pi/(1+r^2)$ and
 $|g^a_2|^2 = 8\pi r^2/(1+r^2)$ at $r=4$.
Along the line of symmetry (purple), $\rho_{12}(\infty)=0$, while
$\Psi_d =0$ defines the black line. The quantum conductances vanish along both lines.
Note that a back flow ($J_d(\infty) <0$) occurs near $\phi^{\rm h}=-\phi^{\rm c} = \pm 1$ in (a),
where the negative quantum current overmatches the positive classical current.
}
\label{fig:2}
\end{figure}

As $\rho_{12}(\infty)$ is also proportional to bias $\Delta N$, the QHE
can be viewed as an analogue of an electronic circuit with resistors
$R_1$ and $R_2$
in parallel under the external potential bias (see the inset of Fig.~\ref{fig:1}).
The conductance $\sigma_d$ of dot $d$
is defined by the Ohm's law of $J_d(\infty)=\sigma_d \Delta N $,
which is the reciprocal of resistance as $\sigma_{d} = R_{d}^{-1}$.
The conductance is also divided into the classical and quantum part as
$\sigma_{d} = \sigma_{d}^{\rm cl} +\sigma_{d}^{\rm q}$ from Eq.~\eqref{eq:Jd}.
The classical part $\sigma_{d}^{\rm cl}$ is always positive, while the quantum part
can be either positive or negative.
In Fig.~\ref{fig:2}, we plot the relative quantum conductance $\sigma_d^{\rm q}/\sigma_d^{\rm cl}$
in the $(\phi^{\rm c},\phi^{\rm h})$ plane in the $r$-symmetric configuration.
Near but off the symmetric line of $\phi^{\rm h}=\phi^{\rm c}$, we find the total quantum conductance
$\sigma^{\rm q}=\sum_d \sigma_d^{\rm q}>0$, which means
that the performance of the QHE can be enhanced beyond the classical limit in this parameter regime.

For small $\Delta N$, we expand the relative quantum
conductance as
\begin{align}
\sigma^{\rm q}_{d}/\sigma^{\rm cl}_{d} = \mathcal{S}_{d}^0 +
\mathcal{S}_{d}^1~\Delta N + \cdots,
\end{align}
 where
$\mathcal{S}_{1}^0 \sim - (\phi^{\rm h} - \phi^{\rm c} )^2$,
$\mathcal{S}_{2}^0 = \mathcal{S}^0_1/r^2$, and
$\mathcal{S}^1_{d} \sim \phi^{\rm h} (\phi^{\rm h} - \phi^{\rm c})$
for the $r$-symmetric configuration
(see Sec.~\ref{sec:current} of SM~\cite{supp} for detailed calculations).
Interestingly, $\sigma^{\rm q}_d$ is always non-positive
in the linear response regime ($\mathcal{S}_{d}^0 \leq 0$), but may become positive
due to $\mathcal{S}_{d}^1$ in the nonlinear regime as $\Delta N$ increases for $\phi^{\rm h}(\phi^{\rm h}-\phi^{\rm c})>0$.
Note that $\mathcal{S}_{d}^1$ can dominate over $\mathcal{S}_{d}^0$ near the symmetric line ($\phi^{\rm h}=\phi^{\rm c}$).
For $r>1$, the negative quantum effect ($\mathcal{S}_{d}^0$) is relatively stronger for dot 1 which has a weaker coupling with baths,
as also seen in Fig.~\ref{fig:2}, which might be applicable to a filtering circuit.

Although $\rho_{12}(\infty)$ becomes finite off the symmetric line ($\phi^{\rm h} \neq \phi^{\rm c}$),
the quantum current may vanish again when $\Psi_d=0$ in Eq.~\eqref{eq:Jd}, which is denoted by
black lines in Fig.~\ref{fig:2}. This can happen by balancing the quantum contributions from the stochastic
part and the interference part, which are represented by first two terms and the third term in the right-hand-side of Eq.~\eqref{eq:Jda},
respectively.
The quantum enhancement occurs only between two lines of $\Psi_d=0$ and $\rho_{12}(\infty)=0$.
For general cases outside of the $r$-symmetric configuration, these two lines are simply tilted
(see Fig.~\ref{fig:s1} in SM~\cite{supp}), but the general features of the QHE are essentially unchanged.

\emph{Coupling-configuration symmetric case} --
We focus on the most symmetric case with $\phi^{\rm h}=\phi^{\rm c}= \phi$ in the $r$-symmetric configuration,
where we find the simple relations that $W_2 = r^2 W_1$, $\overline{W}_2 = r^2 \overline{W}_1$,
$\Phi = r \phi W_1$, and  $\overline{\Phi} = r \phi \overline{W}_1$,
 yielding $W_d/\overline{W}_d=\Phi/\overline{\Phi}$.
Then, the QME in Eq.~\eqref{eq:lind2} can be reduced to the QME with a single {\em effective} bath,
defined by a single coherence parameter $\phi$ and a single rate $W_1$. As is well known for the QME with a single bath,
the system should reach a classical equilibrium state in the long-time limit.
However, with degenerate energy levels, the off-diagonal (coherent) terms in the dissipation
matrix $\Gamma$ in Eq.~\eqref{eq:gamma} cannot be ignored even under the rotational wave approximation.
It turns out that these coherent terms slow down the quantum dynamics significantly in general ($|\phi|<1$), approaching
the classical steady state via a long lived quasi-stationary state with nonzero coherence.

First, we calculate the eigenvectors $\bf{v}_i$ and the corresponding
eigenvalues $\lambda_j$ of the Liouville matrix $\mathsf{L}$.
Details are given in  Sec.~\ref{sec:singular} of SM~\cite{supp}.
We find the steady-state eigenvector
${\bf v}_1^{\rT} = (\bar{\alpha}, \alpha, \alpha, 0,0)$ with $\lambda_1=0$,
where $\alpha = {W_1}/{(2 W_1 + \overline{W}_1)}$ and
$\bar{\alpha} = 1-2\alpha$,
which corresponds to the classical fixed point.
Other eigenvalues are negative except for $|\phi|=1$, thus the classical fixed point
represents the unique steady state. At $|\phi|=1$, however,
another eigenvector $\bf{v}_4$ 
also has the zero eigenvalue, allowing multiple fixed points spanned by
${\bf v}_1$ and ${\bf v}_4$. Note that
$|\mathsf{L}_{\rm ss}|=r^2(1-\phi^2) (2W_1+\overline{W}_1)\overline{W}_1$ from Eq.~\eqref{eq:ss_matrix}, which vanishes at
these singular points of $|\phi|=1$.

\begin{figure}
\includegraphics[width=0.8\columnwidth]{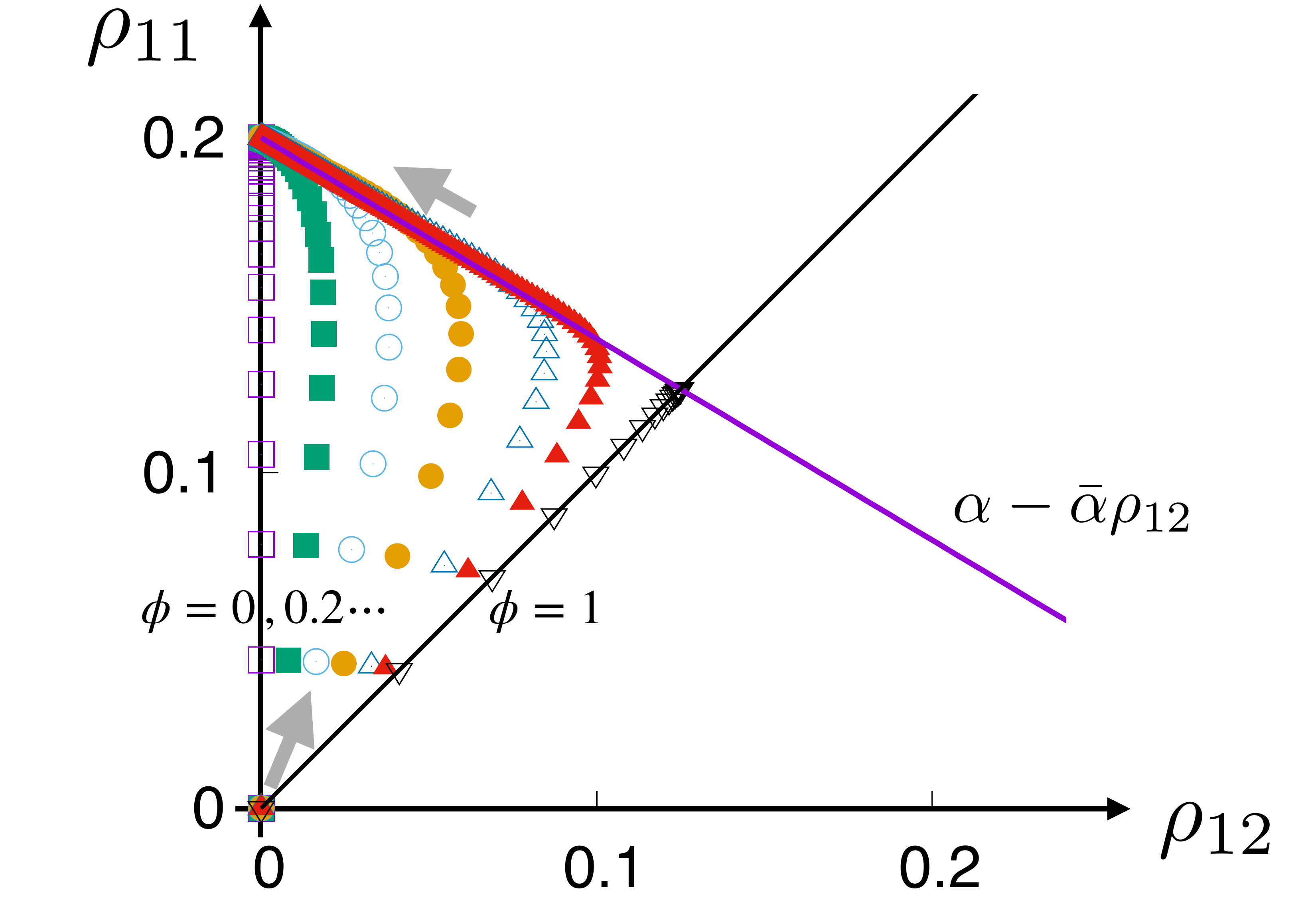}
\caption{ Dynamic trajectories starting from $(\rho_{12},\rho_{11})=(0,0)$
for various $\phi$ with $r=1$, $W_1 =0.25$, and $\overline{W}_1 = 0.75$,
yielding $\alpha=0.2$ and $\overline{\alpha}=0.6$.
Numerical data are denoted
by various symbols for $\phi=0, 0.2, 0.4, 0.6, 0.8, 0.9, 1$ (from left to right).
The time interval between the same symbols is set to be 0.2 and
the grey arrows denote the direction of dynamics.
The classical fixed point is at
$(\rho_{12},\rho_{11})=(0,0.2)$, while the coherent fixed point is
at $(0.125,0.125)$.
}
\label{fig:3}
\end{figure}

Defining a matrix $\sf V = ( {\bf v}_1,  {\bf v}_2,  {\bf v}_3,  {\bf v}_4,  {\bf v}_5)$
with the probability conservation,
the formal solution for $\mathbf{P}(t)$ reads
\begin{equation}
\label{eq:evolv2}
\mathbf{P}(t) = {\sf V} \left( 1, ~\chi_2 e^{\lambda_2 t},
 ~\chi_3 e^{\lambda_3 t},  ~\chi_4 e^{\lambda_4 t},  ~\chi_5 e^{\lambda_5 t} \right)^{\rT} \,,
\end{equation}
where $\chi_i$ depends on the initial condition $\mathbf{P}(0)$ in general.
At $|\phi|=1$, $\lambda_1=\lambda_4=0$, so the steady state $\mathbf{P}(\infty)$
still depends on $\mathbf{P}(0)$. In Fig.~\ref{fig:3}, we display typical dynamic trajectories
in the $(\rho_{12},\rho_{11})$ space with $r=1$,
starting from the empty initial condition of $\rho_{ij}(0)=0$ except for $\rho_{00}(0)=1$.
As expected, all trajectories end up in the single (classical) fixed point in the long-time limit
except for $|\phi|= 1$, where the new coherent fixed point appears with $\rho_{12}(\infty)\neq 0$.
Note that the dynamics for $\phi\lesssim 1$ detours around the coherent fixed point for a significantly long
time (known as a quasi-stationary state before), approaching the classical fixed point, thus it may be observed experimentally as
a {\em quasi}-stationary state even in the presence of
small decoherence.

The additional zero eigenvalue ($\lambda_4=0$) at the singular points ($|\phi|=1$) implies another conservation law beside the probability conservation.
Specifically, we find $r^2\dot{\rho}_{11} + \dot{\rho}_{22} - r\dot{\rho}_{12} - r\dot{\rho}_{21}=0$ for $\phi=1$
from Eq.~\eqref{eq:LW}, or $r^2\rho_{11}(t) + \rho_{22}(t) - r\rho_{12}(t) - r\rho_{21}(t)=I_0$ for all time $t$,
where $I_0$ is a constant determined by the initial condition.
We obtain the steady state solutions using Eq.~\eqref{eq:sol_ss1} and the conservation law, written as
$\rho_{11}(\infty) = \alpha -
[r \bar{\alpha} -\frac{1-r^2}{r}\alpha]\, \rho_{12}(\infty)$ and
$\rho_{22}(\infty) = \alpha -[ \frac{\bar{\alpha}}{r} + \frac{1-r^2}{r} \alpha ]\, \rho_{12}(\infty)$ with
\begin{equation} \label{eq:steady_single_asym}
\rho_{12}(\infty) =\rho_{21}(\infty)=  \frac{r }{1+r^2}
\frac{1}{1-\alpha}
\left(  \alpha - \frac{I_0}{1 + r^2} \right)~,
\end{equation}
which depends on the initial state.  In Fig.~\ref{fig:3}, we set $r=1$ and $I_0=0$, so
the coherent fixed point is determined by the intersection of two lines,
$\rho_{11} = \rho_{12}$ and $\rho_{11} = \alpha - \bar{\alpha} \rho_{12}$.
For $I_0 \neq 0$, the coherent fixed point is shifted along the curve of
$\rho_{11} = \alpha - \bar{\alpha} \rho_{12}$.
In case of $\phi=-1$, we get the same results except for
changing the signs of $\rho_{12}$ and $\rho_{21}$ (see Eq.~\eqref{eq:phi-1} of SM~\cite{supp}).
Note that the coherence can be finite and initial-state dependent even for $\Delta N=0$ (equilibrium).
This may raise a doubt that the quantum current $J_d^{\rm q}$ might not vanish in equilibrium.
Nevertheless $J_d^{\rm q}$ always vanishes in equilibrium as well as the classical current
$J_d^{\rm cl}$, because the quantum speed
$\Psi_d$ is proportional to bias $\Delta N$ (in fact, $\Psi_d=\phi(\frac{1+r^2}{r}) J_d^{\rm cl}$ in Eq.~\eqref{eq:Psi} of SM~\cite{supp}).
The relative quantum conductance can be positive even in the linear response regime,
i.e.~$\mathcal{S}^0_{d}$ can be positive, depending on the initial state.

The phenomena of multiple fixed points are observed not only in
fermionic systems~\cite{schaller}, but also in bosonic systems~\cite{darkstate, juzar}
which result from the existence of a {\em dark} state. In our case, the system state
can be rewritten in a rotated orthonormal bases as $|0\rangle$, $|+\rangle=(|1\rangle+r|2\rangle)/{\mathcal{N}_r}$,
and $|-\rangle=({r|1\rangle-|2\rangle})/{\mathcal{N}_r}$ with
$\mathcal{N}_r= \sqrt{1+r^2}$. Then, the system Hamiltonian is given as
$\hat{H}_{\rm S}=E(|+\rangle\langle+|+|-\rangle\langle-|)$
and the interaction
Hamiltonian becomes $\hat{H}^a_{\rm SB} = {\mathcal{N}_r}\sum_{k} g^a_1 {\hat{b}_k^{a\dagger}}
~|0 \rangle \langle +|+ h.c.$ at the singular points. Note that  the state $|-\rangle$
remains unchanged under the evolution operator,
which corresponds to the dark state at $\phi=1$, i.e.~any initial population in the dark state remains intact or
$\langle-|\hat{\rho}_{\rm S}|-\rangle= (r^2 \rho_{11} + \rho_{22} - r \rho_{12} -r \rho_{21})/\mathcal{N}_r^2$ should be conserved.
We can easily extend our result to the degenerate multiple dots with multiple occupancy allowed.
As the dark state decouples with baths, it may be useful to
protect quantum information from decoherence~\cite{zanner}.

We remark that the  Lindblad description of degenerate quantum dots coupled to a single bath
also yields multiple steady states with coherence at the maximum interference. This might be
against the conventional wisdom that a system coupled to a single bath should reach
the incoherent thermal equilibrium, regardless of its initial state.
In this sense, the phenomenological introduction of $\phi$ is natural to guarantee the
thermal steady state for $|\phi|<1$. Near the singular points, one may observe a long-living
quasi-stationary state with the information of initial-state dependent coherent solutions.

\emph{Conclusion} --
We have investigated all possible steady-state
solutions in the continuous quantum-dot heat engine coupled to terminals in parallel
for various tunneling coefficients and interference strengths.
The interference strength used in this work plays a similar role of
the alignment of dipoles~\cite{dipole} in the bosonic system and
acts as a source for decoherence from environments.
We found that, unless the interference is completely negated, the steady states possess the coherence in general,
which generates an extra quantum current, thus the engine performance can be enhanced
in a specific region of the parameter space.
Recently, the single quantum-dot (fermion) heat engine was realized experimentally~\cite{josefsson}.
The parallel-double-dot engine is also expected to be synthesized to confirm
the enhancement of the QHE performance by thermal noises.

\begin{acknowledgments}
This research was supported by
the NRF Grant No.~2020R1I1A1A01071924 (JU) and No.~2017R1D1A1B06035497 (HP), and
by the KIAS individual Grant No.~PG013604 (HP).
K.E.D. is supported by the National
Science Foundation of China (No. 11934011), the Zijiang
Endowed Young Scholar Fund, the East China Normal University
and the Overseas Expertise Introduction Project for
Discipline Innovation (111 Project, B12024).
\end{acknowledgments}

\end{document}


\title{Supplementary Material for \\
``Coherence enhanced quantum-dot heat engine''}
\author{Jaegon Um}
\email{slung@postech.ac.kr}
\affiliation{Department of Physics, Pohang University of Science and Technology, Pohang 37673, Korea}

\author{Konstantin E.~Dorfman}
\email{dorfmank@lps.ecnu.edu.cn}
\affiliation{State Key Laboratory of Precision Spectroscopy, East China Normal University,
Shanghai 200062, China}

\author{Hyunggyu Park}
\email{hgpark@kias.re.kr}
\affiliation{School of Physics, Korea Institute for Advanced Study, Seoul 02455, Korea}
\affiliation{Quantum Universe Center, Korea Institute for Advanced Study, Seoul 02455, Korea}

\newcommand{\red}[1]{{\color{red}#1}}

\maketitle

\section{Quantum Master Equation}
\label{sec:ME}
\subsection{Model} \label{sec:model}
We start with the Hamiltonian for the system (quantum dots) interacting with
heat bathes, which are given by
\begin{equation}
\hat{H} = \hat{H}_{\rm S} + \hat{H}_{\rm B} + \hat{H}_{\rm SB} \,,
\label{eq:H}
\end{equation}
where $\hat{H}_{\rm S}$, $\hat{H}_{\rm B}$, and
$\hat{H}_{\rm SB}$ denote the Hamiltonians for
the quantum-dot system, heat baths
and interactions between the system and baths,
respectively.
The Hamiltonian of double quantum dots is given by
\begin{equation}
\hat{H}_{\rm S} = \hat{d}^{\dagger}_1 \hat{d}_1
+  E_2 \hat{d}^{\dagger}_2 \hat{d}_2 +
E_{12} \hat{d}^{\dagger}_1 \hat{d}_1 \hat{d}^{\dagger}_2 \hat{d}_2 \,,
\label{eq:Hd}
\end{equation}
where
$E_1$ and $E_2$ denote energies
for dot 1 and dot 2, respectively,
and $E_{12}$ is the Coulomb repulsion between electrons at dots.
In the case of the degenerated dots, $E_1=E_2\equiv E$.
Here $\hat{d}_d$
($\hat{d}^{\dagger}_d$) is the fermionic operator
annihilating (creating) a single particle at dot $d$.
We assume that only a single spinless fermion is allowed for each dot.
Note that coherent hoppings between dots are not allowed.

The bath Hamiltonian is the sum of each bath Hamiltonian $\hat{H}_{\rm B}^a$ of bath $a$ , which can be also written in terms of
fermionic operators as
\begin{equation} \label{eq:Hb}
\hat{H}_{\rm B} =\sum_{a={\rm h,c}}\hat{H}_{\rm B}^a = \sum_{a={\rm h,c}}\sum_{k}
\omega^{a}_{k}\, \hat{b}^{a\dagger}_k  \hat{b}^{a}_k\,,
\end{equation}
where $\hat{b}^{a}_k$ ($\hat{b}^{a\dagger}_k $)
denotes the
operator annihilating (creating) a particle with momentum $k$ and energy
$\omega^{a}_{k}$ in bath $a$ (for simplicity,
we assume here that the momentum is a scalar variable).
The interaction Hamiltonian $H_{\rm SB}$ is also the simple sum of the interaction Hamiltonian $\hat{H}_{\rm SB}^a$ for each bath $a$, which is also expressed with the fermionic operators as
\begin{equation} \label{eq:Hi}
\hat{H}_{\rm SB} = \sum_{a={\rm h,c}} \hat{H}_{\rm SB}^a =\sum_{a={\rm h,c}}\sum_{d=1,2} \sum_{k} g^{ a}_{d}\, \hat{b}^{a\dagger}_{k}
\hat{d}_d \, + h.c. \,,
\end{equation}
which describes an electron hopping between dot $d$ and bath $a$ with
a coupling strength $g^a_d$, usually depending on the
momentum or energy.

In the limit of $E_{12} \to \infty$ (infinite repulsion),
the simultaneous occupation at both dots is prohibited, thus the system state can be
described with the three orthonormal bases of
$|0\rangle$ (empty), $|1\rangle$ (single occupation in dot 1),
and $|2\rangle$ (single occupation in dot 2).
Then,  the operator $\hat{d}_d$ at dot $d$ can be
replaced by a jump operator $|0\rangle\langle d|$.
Using these bases, we rewrite
\begin{equation}
\hat{H}_{\rm S} = E_0 |0\rangle \langle 0| + E_1 |1\rangle \langle 1|
 + E_2 |2\rangle \langle 2|~,\quad \hat{H}_{\rm SB}=\sum_{a} \sum_{d,k} g^{a}_{d}\hat{b}^{a\dagger}_{k}|0\rangle\langle d| + h.c.
\label{eq:Hd_1}
\end{equation}
where $E_0$ means the empty-state energy (here, we set $E_0=0$).

\subsection{Derivation of the QME}

We derive the QME with an assumption that dots and
baths are weakly coupled~\cite{bre02}. Instead of exploiting fermionic operators of dots, used in
previous works~\cite{harbola, schaller, cuetara}, we use the jump operators.
Starting from the von Neumann equation of the {total} system,
$\partial_t \hat{\rho} = -{\rm i} \left[ \hat{H}, \hat{\rho} \right]$,
where $\hat{\rho}(t)$ is the density operator in the Schr{\"o}dinger picture,
the  system dynamics expressed by the reduced density operator, $\hat{\rho}_{\rm S} = {\rm tr}_{\rm B} \,\hat{\rho}$,
is obtained by tracing out the bath degrees of freedom in the total system dynamic equation.
In the weak coupling limit where the interaction Hamiltonian is
small in comparison to the system and bath Hamiltonian,
it is convenient to take the interaction picture, where
$\hat{\rho}'(t) = e^{{\rm i}(\hat{H}_{\rm S}+\hat{H}_{\rm B})t}
\hat{\rho}(t) e^{-{\rm i}(\hat{H}_{\rm S}+\hat{H}_{\rm B})t}$
with $\hat{\rho}(t) = e^{-{\rm i}\hat{H}t} \hat{\rho}(0)e^{i\hat{H}t}$.
Then, the von Neumann equation in the interaction picture becomes
\begin{equation}
\partial_t \hat{\rho}' = -{\rm i} \left[ \hat{H}'_{\rm SB}, \hat{\rho}' \right]\,,
\end{equation}
where the interaction Hamiltonian in the interaction picture
$\hat{H}'_{\rm SB} =
e^{{\rm i}(\hat{H}_{\rm S}+\hat{H}_{\rm B})t}
\hat{H}_{\rm SB} e^{-{\rm i}(\hat{H}_{\rm S}+\hat{H}_{\rm B})t}$ is obtained from Eq.~\eqref{eq:Hd_1} as
\begin{align}
\label{eq:int}
\hat{H}'_{\rm SB} = \sum_{a}\sum_{d,k} \left[ g^a_d \,
e^{{\rm i}\hat{H}_{\rm B}t}\, \hat{b}^{a\dagger}_k e^{-{\rm i}\hat{H}_{\rm B}t}\,
|0 \rangle \langle d|\, e^{-{\rm i} E_d t} + h.c. \right] \,.
\end{align}
Using a formal solution, $\hat{\rho}'(t) = \hat{\rho}(0) -{\rm i}\int_0^t\,
d\tau [ \hat{H}'_{\rm SB}(\tau), \hat{\rho}'(\tau) ] $,
the equation for
$\hat{\rho}'_{\rm S} = {\rm tr}_{\rm B} \,\hat{\rho}'$ is
written as
\begin{equation}
\partial_t \hat{\rho}'_{\rm S}(t) =
-{\rm tr}_{\rm B} \int_0^t d\tau \,\left[ \hat{H}'_{\rm SB}(t) , \left [
\hat{H}'_{\rm SB}(\tau), \hat{\rho}'(\tau) \right ] \right],
\end{equation}
where the initial condition satisfies ${\rm tr}_{\rm B}
[ \hat{H}'_{\rm SB}, \hat{\rho}'(0)]=0$.
Substituting $\tau = t-s$, we obtain
\begin{equation}
\partial_t \hat{\rho}'_{\rm S}(t) = -{\rm tr}_{\rm B}
\int_0^t ds \,\left[ \hat{H}'_{\rm SB}(t) , \left [
\hat{H}'_{\rm SB}(t-s), \hat{\rho}'(t-s) \right ] \right].
\label{eq:step0}
\end{equation}

Now we take the so-called Born-Markov approximation, where
it is assumed that $\hat{\rho}'(t) \approx \hat{\rho}'_{\rm S}(t)
\otimes \hat{\rho}_{\rm B}$
with the canonical heat bath density operator,
\begin{equation}
\hat{\rho}_{\rm B} = Z^{-1}  e^{-\sum_a
\left( \hat{H}^a_{\rm B}-\mu_a \hat{n}_a \right)/T_a } \,,
\label{eq:rhoB}
\end{equation}
with the temperature $T_a$, the chemical potential $\mu_a$, and
the number operator $\hat{n}_a = \sum_{k}  \hat{b}^{a\dagger}_k \hat{b}^a_k$ for each bath $a$,
and the partition function $Z= {\rm tr}_{\rm B} e^{-\sum_a \left( \hat{H}^a_{\rm B}-\mu_a \hat{n}_a \right)/T_a }$ (the Boltzmann constant is set as $k_{\rm B}=1$). As the total density operator is given in the product form,
this assumption implies that each bath is always in its equilibrium, regardless of the system evolution.
This happens when the bath time scale $\tau_{\rm B}^a$ is much smaller than the system time scale, thus
the time scale separation between the system and baths is taken for granted, leading to the approximate
replacement of  $\hat{\rho}'_{\rm S}(t-s) \to \hat{\rho}'_{\rm S}(t)$.
Since the correlation ${\rm tr}_{\rm B} \left[ \hat{H}'_{\rm SB}(t) ,
\left [ \hat{H}'_{\rm SB}(t-s), \hat{\rho}'(t) \right ] \right] $
in Eq.~\eqref{eq:step0} may vanish for $s \gg \tau_{\rm B}^a$,
the integral upper bound can be extended to $\infty$, yielding
a simpler approximate dynamic equation as
\begin{equation}
\partial_t \hat{\rho}'_{\rm S}(t) =
-{\rm tr}_{\rm B} \int_0^{\infty} ds \,\left[ \hat{H}'_{\rm SB}(t) , \left [
\hat{H}'_{\rm SB}(t-s), \hat{\rho}'_{\rm S}(t) \otimes \hat{\rho}_B \right ] \right].
\label{eq:step1}
\end{equation}

Inserting Eq.~\eqref{eq:int} into Eq.\eqref{eq:step1}, one can write each term
in the commutation in Eq.~\eqref{eq:step1} as
\begin{eqnarray}
\label{eq:c1}
&&{\rm tr}_{\rm B} \hat{H}'_{\rm SB}(t) \hat{H}'_{\rm SB}(t-s)
\hat{\rho}'_{\rm S}(t)\otimes \hat{\rho}_{\rm B}
=\sum_{a,k } \left[ \sum_d \left[\lvert g^a_d \rvert^2\ \left( C^a_k(s)\, e^{-{\rm i} E_d s }\, |0 \rangle \langle 0|
+  D^a_k(s)\, e^{{\rm i} E_d s }\, |d \rangle \langle d|\right)\right] \right.\nonumber\\
&&\qquad\qquad\qquad+ \left.   g^{a*}_1 g^a_2 ~D^a_k(s)\, e^{{\rm i}(E_1-E_2)t} e^{{\rm i} E_2s }\, |1 \rangle \langle 2|
+  g^a_1 g^{a*}_2 ~D^a_k(s)\, e^{-{\rm i}(E_1-E_2)t} e^{{\rm i} E_1 s }\, |2 \rangle \langle 1| \right]\,\hat{\rho}'_{\rm S}\,,
\\
\label{eq:c2}
&&{\rm tr}_{\rm B} \hat{H}'_{\rm SB} (t) \hat{\rho}'_{\rm S}(t)
\otimes \hat{\rho}_{\rm B} \hat{H}'_{\rm SB}(t-s)
=\sum_{a,k } \left[ \sum_d \left[\lvert g^a_d \rvert^2 \left(D^a_k(-s) e^{-{\rm i} E_d s }\, |0 \rangle \langle d|
\hat{\rho'}_{\rm S} |d\rangle \langle 0| + C^a_k(-s) e^{{\rm i} E_d s }\, |d \rangle \langle 0|
\hat{\rho}'_{\rm S} |0\rangle \langle d| \right)\right]
\right. \nonumber\\
&&\qquad\qquad+\left.g^{a*}_1 g^a_2 \left( D^a_k(-s) e^{{\rm i}(E_1-E_2)t}
e^{-{\rm i} E_1s } \,|0 \rangle \langle 2| \hat{\rho}'_{\rm S} |1\rangle \langle 0|
+ C^a_k(-s)  e^{{\rm i}(E_1-E_2)t}
e^{{\rm i} E_2s } \,|1 \rangle \langle 0| \hat{\rho}'_{\rm S}|0\rangle \langle 2|\right) \right. \nonumber\\
&&\qquad\qquad+\left.g^a_1 g^{a*}_2 \left( D^a_k(-s) e^{-{\rm i}(E_1-E_2)t}
e^{-{\rm i} E_2s } \,|0 \rangle \langle 1|\hat{\rho}'_{\rm S} |2\rangle \langle 0|
+ C^a_k(-s)  e^{{\rm i}(E_1-E_2)t}
e^{{\rm i} E_2s } \,|2 \rangle \langle 0| \hat{\rho}'_{\rm S}|0\rangle \langle 1|
\right)\right] \,,
\end{eqnarray}
and the remainders are the Hermitian conjugates of Eqs.~\eqref{eq:c1} and \eqref{eq:c2}.
Note that each bath contributes additively to Eq.~\eqref{eq:step1}.
and the correlators for bath $a$ are defined by
\begin{align}
\label{eq:ck1}
C^a_k(s)
= {\rm tr}_{\rm B} ~e^{{\rm i}\hat{H}_{\rm B} s}
\hat{b}^{a\dagger}_k  e^{-{\rm i}\hat{H}_{\rm B} s}
\, \hat{b}^a_{k}\, \hat{\rho}_{\rm B}
 D^a_k(s)
= {\rm tr}_{\rm B} ~e^{{\rm i}\hat{H}_{\rm B} s}
\, \hat{b}^a_{k}\, e^{-{\rm i}\hat{H}_{\rm B} s}
\hat{b}^{a\dagger}_k  \hat{\rho}_{\rm B}
\,.
\end{align}
With the Fock-state description of bath particles in Eq.~\eqref{eq:rhoB},
we find
\begin{equation}
C^a_k(s) =  N^a(\omega^a_k) e^{{\rm i} \omega^a_k s} \quad {\rm and}\quad
D^a_k(s) = \left[ 1- N^a(\omega^a_k ) \right] e^{-{\rm i} \omega^a_k s}\,
\label{eq:thermal}
\end{equation}
where $N_a$ is the
Fermi-Dirac distribution in bath $a$, given as
\begin{equation}
N^a = \frac{\exp[-(\omega-\mu_a)/T_a ]}{ 1+ \exp[-(\omega-\mu_a)/T_a ]  }.
\end{equation}
Since the integral over time $s$ in Eq.~\eqref{eq:step1} yields the delta function, i.e.,
\begin{equation}
\int_0^{\infty} ds e^{\pm {\rm i}(\omega^a_k- E_d)s} = \pi \delta(\omega^a_k -  E_d)~,
\end{equation}
a single mode for each bath satisfying $\omega^a_{k}=E$
survives in Eq.~\eqref{eq:step1} for the degenerate case with $E_1=E_2=E$.
Note that we have omitted the Lamb shift correction, which is the order of $E^{-1}$,
negligible in the high energy limit.

Changing $\sum_{k} \to \mathcal{N} \int dk$ with a proper normalization $\mathcal{N}$
and integrating over $k$, we calculate the transition rates. First, consider the incoherent terms such as
$|d\rangle \langle 0| \hat{\rho}'_{\rm S} |0 \rangle \langle d|$ and
$|0\rangle \langle d| \hat{\rho}'_{\rm S} |d \rangle \langle 0|$.
For transitions between $|0\rangle$ and $|d\rangle$ due to bath  $a$,
the transitions rates are obtained as
\begin{equation}
\label{eq:rate1}
w^a_{d +} = 2\pi \lvert g^a_d \rvert^2 N^a( E)\quad {\rm and}\quad
w^a_{d -} = 2\pi \lvert g^a_d \rvert^2 \overline{N^a}(E) \,,
\end{equation}
where $\overline{N^a} = 1 -N^a$ and $g^a_d = g^a_d(E)$.
The $+$ sign in Eq.~\eqref{eq:rate1} stands for the transition
from $|0\rangle$ to $|d\rangle$ and the $-$ sign stands for the opposite direction.
Now we consider interference terms such as
$|0 \rangle \langle 2| \hat{\rho}'_{\rm S} |1 \rangle \langle 0| $ or
$|1\rangle \langle 0| \hat{\rho}'_{\rm S} |0 \rangle \langle 2|$.
Due to the phase factor $\exp[\pm {\rm i} (E_1 -E_2)t]$,
the interference terms vanish in long-time limit unless $E_1=E_2$ (rotational wave approximation).
In this work with $E_1 = E_2=E$, we find the  nonvanishing interference terms as
\begin{equation}
\label{eq:rate3}
\sqrt{w^a_{1 +} w^a_{2 +}} e^{ \pm {\rm i}\theta^a} \quad {\rm and}\quad
\sqrt{w^a_{1 -} w^a_{2 -}} e^{\pm {\rm i}\theta^a} \,,
\end{equation}
where $\theta^a$ is the difference of phase angles between $g^a_1$ and $g^a_2$,
defined as $g^{a*}_1  g^a_2
=\lvert g^a_1 \rvert \lvert g^a_2 \rvert e^{{\rm i} \theta^a}$.

Defining the Lindblad operators as
\begin{equation}
\label{eq:lo}
\hat{L}_{1} =  |1 \rangle \langle 0|\,, \quad \hat{L}_{2} =  |2 \rangle \langle 0|\,,
\quad \hat{L}_{3} =  |0 \rangle \langle 1|\,, \quad
\hat{L}_{4} =  |0 \rangle \langle 2|\,,
\end{equation}
the dynamic equation for the density operator in Eq.~\eqref{eq:step1} is rewritten as
\begin{eqnarray}
\label{eq:lind}
\partial_t \hat{\rho}'_{\rm S} &=& \sum_{a} \left[
w^a_{1+}\left( \hat{L}_{1 } \hat{\rho}'_{\rm S} \hat{L}^{\dagger}_{1 }
-\frac{1}{2} \left\{ \hat{L}^{\dagger}_{1 } \hat{L}_{1 }, \hat{\rho}'_{\rm S}
\right\} \right) + w^a_{2+}\left( \hat{L}_{2 } \hat{\rho}'_{\rm S} \hat{L}^{\dagger}_{2 }
-\frac{1}{2} \left\{ \hat{L}^{\dagger}_{2 } \hat{L}_{2 }, \hat{\rho}'_{\rm S}
\right\}\right)\right. \nonumber\\
&&+w^a_{1-} \left( \hat{L}_{3} \hat{\rho}'_{\rm S} \hat{L}^{\dagger}_{3}
-\frac{1}{2} \left\{ \hat{L}^{\dagger}_{3} \hat{L}_{3} ,\hat{\rho}'_{\rm S}
\right \} \right) + w^a_{2-} \left( \hat{L}_{4} \hat{\rho}'_{\rm S} \hat{L}^{\dagger}_{4}
-\frac{1}{2} \left\{ \hat{L}^{\dagger}_{4} \hat{L}_{4} ,\hat{\rho}'_{\rm S}
\right \} \right) \nonumber\\
&& + \sqrt{w^a_{1+} w^a_{2+}}  e^{{\rm i}\theta^a}  \hat{L}_{1} \hat{\rho}'_{\rm S} \hat{L}^{\dagger}_{2 }
+ \sqrt{w^a_{1+} w^a_{2+}} e^{-{\rm i}\theta^a}   \hat{L}_{2 } \hat{\rho}'_{\rm S} \hat{L}^{\dagger}_{1 }
\\
&&\left.+ \sqrt{w^a_{1-} w^a_{2-}}e^{-{\rm i}\theta_a}  \left(\hat{L}_{3} \hat{\rho}'_{\rm S} \hat{L}^{\dagger}_{4 }
-\frac{1}{2} \left\{ \hat{L}^{\dagger}_{4 } \hat{L}_{3 }, \hat{\rho}'_{\rm S}0
 \right\} \right)+ \sqrt{w^a_{1-} w^a_{2-}} e^{{\rm i}\theta_a}  \left( \hat{L}_{4} \hat{\rho}'_{\rm S} \hat{L}^{\dagger}_{3 }
-\frac{1}{2} \left\{ \hat{L}^{\dagger}_{3 } \hat{L}_{4 }, \hat{\rho}'_{\rm S}
 \right\} \right) \right]  \,, \nonumber
\end{eqnarray}
where $\{\, , \}$ denotes the anticommutator.
We introduce a phenomenological prefactor $\phi_a$ for the interference terms
in Eq.~\eqref{eq:lind} by replacing
$e^{{\rm i}\theta^a} \to \phi^a $ with $|\phi^a|\leq 1$
to take into account decoherence effects by other unknown environmental noises~\cite{weng}.
Note that the phase difference $\theta^a$ is absorbed into $\phi^a$.
In the Schr{\"o}dinger picture with $\hat{\rho}_{\rm S}(t) = e^{-{\rm i} \hat{H}_{\rm S} t}
\hat{\rho}'_{\rm S}(t) e^{{\rm i} \hat{H}_{\rm S} t}$, Eq.~\eqref{eq:lind} is rewritten as a matrix form
in Eqs.~\eqref{eq:lind2} and \eqref{eq:gamma}of the main text.

\section{Eigenvectors and eigenvalues of the Liouville operator}\label{sec:eigen}

The density operator can be written in a form of vector:
$\mathbf{P} = \left(\rho_{00}, \rho_{11}, \rho_{22}, \rho_{12}, \rho_{21},
\rho_{01}, \rho_{02}, \rho_{10}, \rho_{20} \right)^{\rT}$, with
$\rho_{ij}=\langle i|\hat{\rho}_{\rm S}|j\rangle$.
Then, the equation of motion is given by
$\partial_t \mathbf{P} =\mathsf{L}^{\rm tot}\, \mathbf{P}$, where
the Liouville operator $\mathsf{L}^{\rm tot}$ has a form of
\begin{equation} \label{eq:lind4}
\mathsf{L}^{\rm tot} = \left( \begin{array}{cc} \mathsf{L} & 0 \\
0& \mathsf{L}_{\rm irr} + \mathsf{E} \end{array} \right)\,,
\end{equation}
where the upper $5 \times 5$ block is given by
$\mathsf{L} = \sum_a \mathsf{L}^a$ and the first term of the lower $4\times 4$ block
$\mathsf{L}_{\rm irr} = \sum_a \mathsf{L}^a_{\rm irr}$.
Each term in the summation is given as
\begin{equation}
\label{eq:La}
\mathsf{L}^a= \left(
\begin{array}{ccccc}
-\left( w^a_{1+} + w^a_{2+}\right) & w^a_{1-} & w^a_{2-} & \phi^{a*}  \sqrt{w^a_{1-}w^a_{2-}}
& \phi^a  \sqrt{w^a_{1-}w^a_{2-}} \\
w^a_{1+} & -w^a_{1-} & 0 & -\phi^{a*}   \sqrt{w^a_{1-} w^a_{2-}}/2
& -\phi^{a}   \sqrt{w^a_{1-} w^a_{2-}}/2 \\
w^a_{2+} & 0 & -w^a_{2-} & -\phi^{a*}  \sqrt{w^a_{1-} w^a_{2-}}/2
& -\phi^{a}   \sqrt{w^a_{1-} w^a_{2-}}/2 \\
\phi^a  \sqrt{w^a_{1+} w^a_{2+}} &  -\phi^{a} \sqrt{w^a_{1-} w^a_{2-}}/2
& -\phi^a  \sqrt{w^a_{1-} w^a_{2-}}/2 & -\left( w^a_{1-}+w^a_{2-}\right)/2 & 0 \\
\phi^{a*}  \sqrt{w^a_{1+} w^a_{2+}} & -\phi^{a*} \sqrt{w^a_{1-} w^a_{2-}}/2
& -\phi^{a*}  \sqrt{w^a_{1-} w^a_{2-}}/2  & 0 & -\left( w^a_{1-}+w^a_{2-} \right)/2
\end{array} \right)~,
\end{equation}
\begin{equation}
\label{eq:Lirr}
\mathsf{L}^a_{\rm irr}=\left(\begin{array}{cccc}
- \left(w^a_{1+}+w^a_{2+}+w^a_{1-}\right)/2  & -\phi^{a*}  \sqrt{w^a_{1-} w^a_{2-}}/2
& 0 & 0 \\
-\phi^a  \sqrt{w^a_{1-} w^a_{2-}}/2 & -\left( w^a_{1+}+w^a_{2+}+w^a_{2-} \right)/2
& 0 & 0 \\
0 & 0 &-\left(w^a_{1+}+w^a_{2+}+w^a_{1-}\right)/2  &  -\phi^a  \sqrt{w^a_{1-} w^a_{2-}}/2 \\
0 & 0 &-\phi^{a*} \sqrt{w^a_{1-} w^a_{2-}}/2
& -\left(w^a_{1+}+w^a_{2+}+w^a_{2-}\right)/2 \end{array} \right) \,,
\end{equation}
and the second term of the lower block $\mathsf{E}$ is
\begin{equation}
\mathsf{E} = \left(\begin{array}{cccc}
iE & 0 & 0 & 0 \\
0 & i E & 0 & 0 \\
0 & 0 & -i E &  0  \\
0 & 0 & 0 & -i E \end{array} \right) \,.
\end{equation}
It is easy to see that each $2\times 2$ subblock of $\mathsf{L}_{\rm irr}^a$ has negative eigenvalues only
for $|\phi^a|\leq 1$, thus
$\rho_{01}$, $\rho_{02}$, $\rho_{10}$, and $\rho_{20}$, associated with $\mathsf{L}_{\rm irr}$ will
vanish in long-time limit as the pure imaginary $\mathsf{E}$ contributes to a modulation only.

From now on, we focus on the $5\times 5$ matrix $\mathsf{L}$ with
the reduced vector $\mathbf{P} = \left(\rho_{00}, \rho_{11}, \rho_{22}, \rho_{12}, \rho_{21}\right)^{\rT}$,
satisfying the dynamic equation $\partial_t \mathbf{P} =\mathsf{L}\, \mathbf{P}$.
For convenience, we take $\phi^a$ as a real number.
We introduce collective parameters for the sake of brevity as
\begin{eqnarray}
W_{d} = \sum_a w^a_{d+}\,, \quad \overline{W}_{d} = \sum_a w^a_{d-}\,, \quad
\Phi = \sum_a \phi^{a} \sqrt{w^a_{1 +}w^a_{2 +} }\,, \quad
\overline{\Phi} = \sum_a \phi^{a} \sqrt{w^a_{1 -}w^a_{2 -} } \,,
\end{eqnarray}
and then Eq.~\eqref{eq:LW} of the main text is obtained.

From Eq.~\eqref{eq:ss_general}, we find the steady-state solution
by inverting the $2\times 2$ matrix $\mathsf{L}_{\rm ss}$ when its determinant $|\mathsf{L}_{\rm ss}|\neq 0$ as
\begin{eqnarray}
\label{eq:sol_ss3}
\rho_{11}(\infty) &=& \frac{W_1 \overline{W}_2 - \overline{\Phi} \left[
2 \Phi \overline{W}_2 + \overline{\Phi} \left( W_1-W_2 \right) \right] / \left( \overline{W}_1 + \overline{W}_2 \right) }
{| \mathsf{L}_{\rm ss} | } \,, \nonumber\\
\rho_{22}(\infty) &=& \frac{\overline{W}_1 W_2 - \overline{\Phi} \left[
2 \Phi \overline{W}_1 + \overline{\Phi} \left( W_2-W_1 \right) \right] / \left( \overline{W}_1 + \overline{W}_2 \right) }
{| \mathsf{L}_{\rm ss} | } \,, \nonumber\\
\rho_{12}(\infty) &=& \rho_{21}(\infty)=\frac{ 2 \Phi \overline{W}_1 \overline{W}_2 - \overline{\Phi}
\left( W_1 \overline{W}_2 + \overline{W}_1 W_2 \right) }
{ | \mathsf{L}_{\rm ss} |  \left( \overline{W}_1 + \overline{W}_2 \right)  } \,,
\end{eqnarray}
with 
\begin{equation} \label{eq:Lss_det}
|\mathsf{L}_{\rm ss}| =  W_1 \overline{W}_2
+ \overline{W}_1 W_2 + \overline{W}_1 \overline{W}_2 - \overline{\Phi}
\left( 2 \Phi + \overline{\Phi} \right) \,,
\end{equation}
and $\rho_{00}$ can be obtained from the probability conservation of $\rho_{00} = 1- \rho_{11} -\rho_{22}$.
This steady-state solution should correspond to the eigenvector ${\bf v}_1$ of the $\mathsf{L}$ matrix with
the eigenvalue $\lambda_1=0$, where
\begin{equation}
{\bf v}_1 = \left( \rho_{00}(\infty), \rho_{11}(\infty), \rho_{22}(\infty),
\rho_{12}(\infty), \rho_{12}(\infty)  \right)^{\rT} \,.
\end{equation}
The other eigenvectors and eigenvalues are reported as below for completeness.
The two eigenvectors, ${\bf v}_2$ and ${\bf v}_3$, have the degenerate eigenvalues
$\lambda_2=\lambda_3= -\left( \overline{W}_1 + \overline{W}_2 \right)/2$, where
\begin{eqnarray}
{\bf v}_{2} = (0,0,0,1,-1)^\rT \quad {\rm and}\quad
{\bf v}_{3} = \left( 0, 1,-1, \frac{ \overline{W}_2 - \overline{W}_1  }
{2 \bar{\Phi} },
\frac{ \overline{W}_2 - \overline{W}_1  }
{2 \bar{\Phi} }  \right )^\rT \,.
\end{eqnarray}
The eigenvalues of the remaining two eigenvectors are the two roots of the characteristic equation of
$\lambda^2 + \lambda \left( W_1 + \overline{W}_1 + W_2 + \overline{W}_2 \right) + |\mathsf{L}_{\rm ss}| =0$.
Thus, we find the eigenvalues $\lambda_4$ and $\lambda_5$ as
\begin{equation}
\lambda_{4,5} =  \frac{ -\left(W_1 + \overline{W}_1
+ W_2 + \overline{W}_2  \right) \pm U }{2} \,,
\end{equation}
with $U = \sqrt{ \left( W_1 + \overline{W}_1 + W_2 + \overline{W}_2 \right)^2  - 4 |\mathsf{L}_{\rm ss}| }$,
where $\lambda_4$ and $\lambda_5$ correspond to the $+$ and $-$ sign, respectively.
The explicit expressions for ${\bf v}_4$ and ${\bf v}_5$ are shown as
\begin{equation}
{\bf v}_{4(5)} = \left(1, - \frac{ \lambda_{4(5)} + W_2 + \overline{W}_2  -W_1  }
{ 2 \lambda_{4(5)} + \overline{W}_1 + \overline{W}_2  } ,
- \frac{ \lambda_{4(5)}+ W_1 + \overline{W}_1  -  W_2 }
{ 2 \lambda_{4(5)} + \overline{W}_1 + \overline{W}_2  },
\,\frac{ 2 \Phi + \overline{\Phi} } { 2 \lambda_{4(5)} + \overline{W}_1 + \overline{W}_2  }\,,
\,\frac{ 2 \Phi + \overline{\Phi} } { 2 \lambda_{4(5)} + \overline{W}_1 + \overline{W}_2  }
 \right)^\rT \,.
\end{equation}
Note that $|\mathsf{L}_{\rm ss}|=0$ yields
the additional zero eigenvalue, $\lambda_4 =0$ and we expect
multiple steady-state solutions given by a linear combination of ${\bf v}_1$ and ${\bf v}_4$.

\section{Steady-state currents} \label{sec:current}

In this section, we calculate steady-state currents explicitly.
The net particle currents $J_d^a$ from bath $a$ to dot $d$  can be calculated
from the Liouville equation in Eqs.~\eqref{eq:lind3} and \eqref{eq:LW} of the main text by
sorting out the contributions to the time increment of the particle density of each dot $d$
($\dot{\rho}_{dd}$) from each bath $a$. Then, we can easily identify
\begin{eqnarray} \label{eq:current}
J^{a}_{d} (t) = w^{a}_{d+}\, \rho_{00}(t) -
w^{a}_{d-}\, \rho_{dd}(t) - \phi^a \sqrt{w^{a}_{1-}w^{a}_{2-} } \,\left(\frac{\rho_{12}(t)+\rho_{21}(t)}{2}\right) \,
\end{eqnarray}
where the first term in the right-hand-side represents particle transfer from bath $a$ to the
empty dot $d$, the second term represents  particle transfer form the occupied dot $d$ to bath $a$,
and finally the third term represents the interference between relaxation channels to both dots.
In the steady state, the particle density at dots is stationary, so
the currents from both baths should be balanced in such a way that
$J^{\rm h}_d(\infty)=-J^{\rm c}_d(\infty)\equiv J_d(\infty)$.

\begin{figure}
\includegraphics[width=1.0\columnwidth]{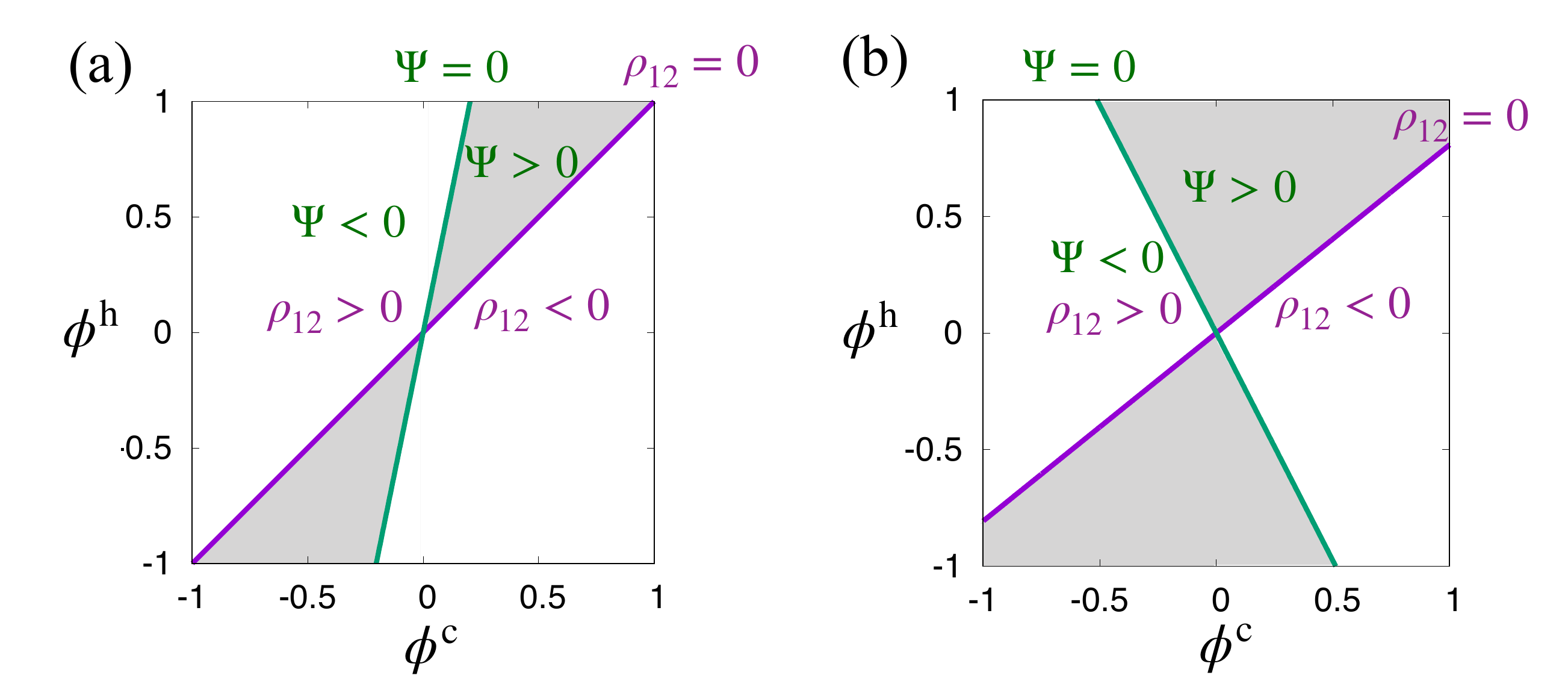}
\caption{
Lines of vanishing quantum currents ($J^{\rm q} = \Psi \,\rho_{12}(\infty) =0$)
in the $(\phi^{\rm c},\phi^{\rm h})$ plane.
The purple line represents $\rho_{12}(\infty)=0$ and the green line represents
$\Psi=0$. The engine performance is enhanced by the extra quantum current in the
shaded regions ($J_{\rm q} > 0$).
We used $N^{\rm h}=0.25$, $N^{\rm c}=0.1$ and tunneling coefficients as
$g^{\rm h}_1 = g^{\rm c}_1 =1/\sqrt{2\pi}$ for dot 1 and (a) $g^{\rm h}_2 = g^{\rm c}_2=4/\sqrt{2\pi}$
($r$-symmetric configuration with $r=4$) and
(b) $g^{\rm h}_2 =12/\sqrt{2\pi}$ and $g^{\rm c}_2=4/\sqrt{2\pi}$ ($r$-symmetry broken) for dot 2.
}
\label{fig:s1}
\end{figure}

We can divide the particle current into the classical and quantum part as
\begin{align}\label{eq:cq_current}
J_d(\infty)=J_d^{\rm cl} + J_d^{\rm q}= J_d^{\rm cl} + \Psi_d \rho_{12}(\infty),
\end{align}
where the quantum current $J_d^{\rm q}$ is given as a product of the \emph{quantum speed} $\Psi_d$ and
 the coherence $\rho_{12}(\infty)$.
The classical current is easily obtained by simply setting
$\phi^a=0$ in Eq.~\eqref{eq:current} and Eq.~\eqref{eq:sol_ss1} of the main text, and using the rates $w^a_{d+} = 2\pi \lvert g^a_{d}\rvert ^2 N^a$ and
 $w^a_{d-}=  2\pi \lvert g^a_{d}\rvert ^2 \overline{{N}^{a}} $, yielding
\begin{align} \label{eq:Jcl1}
&J^{\rm cl}_{1} = \frac{1}{|\mathsf{L}_0|}\left(w_{1+}^{\rm h} \overline{W}_1-w_{1-}^{\rm h} {W}_1\right)\overline{W}_2
= \frac{\Delta N} {|\mathsf{L}_0|} (2\pi)^2
\lvert g^{\rm h}_1 \rvert^2 \lvert g^{\rm c}_1 \rvert^2 \overline{W}_2 \,,\\
\label{eq:Jcl2}
&J^{\rm cl}_{2} = \frac{1}{|\mathsf{L}_0|}\left(w_{2+}^{\rm h} \overline{W}_2-w_{2-}^{\rm h} {W}_2\right)\overline{W}_1
=\frac{\Delta N}  {|\mathsf{L}_0|} (2\pi)^2
\lvert g^{\rm h}_2 \rvert^2 \lvert g^{\rm c}_2 \rvert^2 \overline{W}_1 \,,
\end{align}
where $|\mathsf{L}_0|=|\mathsf{L}_{\rm ss}|_{\phi^a=0}=W_1 \overline{W}_2 + \overline{W}_1 W_2 +  \overline{W}_1  \overline{W}_2$
and the external bias $\Delta N=N^{\rm h}-N^{\rm c}>0$. Note that $J_d^{\rm cl}$ is always positive.
The quantum part, $J^{\rm q}_{d} = \Psi_{d}\, \rho_{12}(\infty)$, is also obtained from Eq.~\eqref{eq:current} and
Eq.~\eqref{eq:sol_ss1} of the main text,
yielding
\begin{align} \label{eq:Psi1}
&\Psi_1 = \frac{\overline{\Phi}}{|\mathsf{L}_0|}
\left[ \left( w_{1+}^{\rm h} \overline{W}_1-w_{1-}^{\rm h} {W}_1 \right) + w^{\rm h}_{1-} W_2 +
\left(w^{\rm h}_{1+}+w^{\rm h}_{1-}\right)\overline{W}_2\right] - \phi^{\rm h} \sqrt{w^{a}_{1-}w^{a}_{2-} }
\,,\\
\label{eq:Psi2}
&\Psi_2 = \frac{\overline{\Phi}}{|\mathsf{L}_0|}
\left[ \left( w_{2+}^{\rm h} \overline{W}_2-w_{2-}^{\rm h} {W}_2 \right) + w^{\rm h}_{2-} W_1 +
\left(w^{\rm h}_{2+}+w^{\rm h}_{2-}\right)\overline{W}_1\right] - \phi^{\rm h} \sqrt{w^{a}_{1-}w^{a}_{2-} } \,,
\end{align}
and the coherence term $\rho_{12}(\infty)=\rho_{21}(\infty)$ is obtained from Eq.~\eqref{eq:sol_ss3}, after some algebra, as
\begin{eqnarray} \label{eq:rho12_full}
\rho_{12}(\infty)= \frac{\Delta N (2\pi)^2 } {|\mathsf{L}_{\rm ss}|}
\frac{|g_1^{\rm h}| |g_2^{\rm h}| \phi^{\rm h}
\left(\lvert g^{\rm c}_2 \rvert^2 \overline{W}_1 +\lvert g^{\rm c}_1 \rvert^2 \overline{W}_2   \right)
- |g_1^{\rm c}|| g_2^{\rm c}| \phi^{\rm c}
\left(\lvert g^{\rm h}_2 \rvert^2 \overline{W}_1 +\lvert g^{\rm h}_1 \rvert^2 \overline{W}_2   \right) }
{\overline{W}_1+\overline{W}_2} \,,
\end{eqnarray}
which is valid except for the singular case of $|\mathsf{L}_{\rm ss}|=0$ (see Sec.~\ref{sec:singular} for the singular case).
In contrast to the classical current, the quantum current $J_d^{\rm q}$ can be both positive and negative,
which can vanish either by the zero quantum speed ($\Psi_d=0$) or by the zero coherence ($\rho_{12}=0$).
In Fig.~\ref{fig:2} of the main text, the lines of $\Psi_d=0$ and
$\rho_{12}(\infty)=0$ are plotted in the $(\phi^{\rm c},\phi^{\rm h})$ plane
for the $r$-symmetric configuration. Note that the $\Psi_d=0$ lines can be different from each other.
For the total quantum current $J^{\rm q}=\sum_d J^{\rm q}_d=\Psi\rho_{12}(\infty)$ with $\Psi=\sum_d \Psi_d$,
the lines of $\Psi=0$ and $\rho_{12}(\infty)=0$ are shown in
Fig.~\ref{fig:s1}, where $J^{\rm q}>0$ is accomplished only in the shaded area with the same signs of
$\Psi$ and $\rho_{12}(\infty)$. Therefore, the engine performance can be enhanced due to the extra positive quantum
current in a specific region of the parameter space.

As can be seen in Eqs.~\eqref{eq:Jcl1}, \eqref{eq:Jcl2}, and \eqref{eq:rho12_full}, both the classical and quantum current are proportional to the external bias $\Delta N$. Thus, it is natural to define the conductance $\sigma_{d}$ for dot $d$
as $J_{d}(\infty) \equiv \sigma_{d} \, \Delta N$, which is also divided into the classical and quantum contribution as
$ \sigma_{d} = \sigma^{\rm cl}_{d} + \sigma^{\rm q}_{d}$. The classical conductance $\sigma^{\rm cl}_{d} $ can be easily obtained from
Eqs.~\eqref{eq:Jcl1} and \eqref{eq:Jcl2}, which is always positive ($\sigma_d^{\rm cl}>0$). The quantum conductance can be obtained from
Eqs.~\eqref{eq:Psi1}, \eqref{eq:Psi2}, and \eqref{eq:rho12_full}. It would be interesting to study the total quantum conductance
$\sigma^{\rm q}=\sum_d \sigma_d^{\rm q}$ in the linear response regime for small bias $\Delta N$.
For convenience, we introduce the mean bias $N\equiv (N^{\rm h}+ N^{\rm c})/2$ with $\overline{N}\equiv 1-N$.
After some algebra, we find 
\begin{equation}
\label{eq:linear_sigma}
\lim_{\Delta N\rightarrow 0} \sigma^{\rm q} =
-\frac{ (2\pi)^2\left[ |g_1^{\rm h}|| g_2^{\rm h}| \phi^{\rm h}
\left(\lvert g^{\rm c}_2 \rvert^2 \overline{W}^{\rm eq}_1 +\lvert g^{\rm c}_1 \rvert^2 \overline{W}^{\rm eq}_2   \right)
- |g_1^{\rm c}||g_2^{\rm c}| \phi^{\rm c}
\left(\lvert g^{\rm h}_2 \rvert^2 \overline{W}^{\rm eq}_1 +\lvert g^{\rm h}_1 \rvert^2 \overline{W}^{\rm eq}_2   \right)
\right]^2 }
{\left( \lvert g^{\rm h}_1 \rvert^2 +  \lvert g^{\rm c}_1 \rvert^2  \right)
\left( \lvert g^{\rm h}_2 \rvert^2 +  \lvert g^{\rm c}_2 \rvert^2  \right)
\left( \overline{W}^{\rm eq}_1 + \overline{W}^{\rm eq}_2 \right) \lvert \mathsf{L}_{\rm ss} \rvert^{\rm eq}  } \,,
\end{equation}
where $\overline{W}^{\rm eq}_d = 2\pi \overline{N} \left( \lvert g^{\rm h}_d \rvert^2 + \lvert g^{\rm c}_d \rvert^2 \right)$
and $| \mathsf{L}_{\rm ss} |^{\rm eq}$ from Eq.~\eqref{eq:Lss_det} as
\begin{eqnarray}
| \mathsf{L}_{\rm ss} |^{\rm eq} = (2\pi)^2 (1+N)\overline{N}
\left[ \left(|g^{\rm h}_1|^2 + |g^{\rm c}_1|^2 \right)
\left( |g^{\rm h}_2|^2 + |g^{\rm c}_2|^2 \right)-\left( |g^{\rm h}_1| |g^{\rm h}_2|\,\phi^{\rm h} +
|g^{\rm c}_1| |g^{\rm c}_2|\,\phi^{\rm c} \right)^2 \right]\,.
\nonumber
\end{eqnarray}
Note that the quantum conductance in Eq.~\eqref{eq:linear_sigma} cannot be positive from ,
implying the current enhancement is not possible in the linear response regime.

Next, we will investigate the nonlinear regime in the $r$-symmetric configuration
($g^a_2=r g^a_1$), where the algebra becomes simplified. From Eqs.~\eqref{eq:Jcl1} and ~\eqref{eq:Jcl2},
we can easily see that the classical currents for two dots are simply related as $J_2^{\rm cl}=r^2 J_1^{\rm cl}$.
For convenience, we set the tunnelling coefficients as
\begin{equation}
\lvert g^{a}_{1} \rvert^2 = \frac{ k^{a}}{1+r^2}~,\quad \lvert g^{a}_{2} \rvert^2 = \frac{r^2 k^{a}}{1+r^2}~,
\label{eq:r-config}
\end{equation}
which satisfies the $r$-symmetric condition.
Then, we find
\begin{equation}
\sigma^{\rm cl}_1 = \frac{2\pi}{1+r^2}\, \frac{k^{\rm h} k^{\rm c} }
{k^{\rm h} \widetilde{N}^{\rm h} + k^{\rm c} \widetilde{N}^{\rm c} } \,, ~~~~
\sigma^{\rm cl}_2 = r^2 \sigma^{\rm cl}_1\,,
\end{equation}
where $\widetilde{N}^a = 1 + N^a$.
After some algebra, we also find the relative quantum conductance $\sigma^{\rm q}_{d}/\sigma^{\rm cl}_{d}$ as
\begin{eqnarray}\label{eq:cond1_rconfig}
\frac{\sigma^{\rm q}_{1}}{\sigma^{\rm cl}_1} &=& \frac{r^2}{1+r^2}
\frac{ 2\left( \phi^{\rm h} -\phi^{\rm c}\right)  }
{\mathcal{L}  } \frac{ k^{\rm h} k^{\rm c} } {k^{\rm h} \overline{N}^{\rm h} + k^{\rm c} \overline{N}^{\rm c}  }
\left[  \phi^{\rm c} \left\{ \left( k^{\rm h} + k^{\rm c} \right)
- \left( k^{\rm h} N^{\rm h} + k^{\rm c} N^{\rm c} \right) N^{\rm h} \right \} \overline{N}^{\rm c} \right.\nonumber\\
&&\qquad\qquad\qquad\qquad\qquad\qquad\qquad\left.- \phi^{\rm h} \left\{ \left( k^{\rm h} + k^{\rm c} \right)
- \left( k^{\rm h} N^{\rm h} + k^{\rm c} N^{\rm c} \right) N^{\rm c} \right \} \overline{N}^{\rm h}
\right] \nonumber\\
&& +\frac{1}{1+r^2}
\frac{ 2\left( \phi^{\rm h} -\phi^{\rm c}\right) }{\mathcal{L} }
\frac{k^{\rm h} k^{\rm c} }
{ k^{\rm h} \overline{N}^{\rm h} + k^{\rm c} \overline{N}^{\rm c} }
 \left( \phi^{\rm h}k^{\rm h} \overline{N}^{\rm h}
+\phi^{\rm c} k^{\rm c} \overline{N}^{\rm c} \right)  \left( N^{\rm h} - N^{\rm c} \right) \,,
\end{eqnarray}
\begin{eqnarray} \label{eq:cond2_rconfig}
\frac{\sigma^{\rm q}_{2}}{\sigma^{\rm cl}_2} &=& \frac{1}{1+r^2}
\frac{ 2\left( \phi^{\rm h} -\phi^{\rm c}\right)}{\mathcal{L}}
\frac{ k^{\rm h} k^{\rm c} }
{ k^{\rm h} \overline{N}^{\rm h} + k^{\rm c} \overline{N}^{\rm c} }
\left[  \phi^{\rm c} \left\{ \left( k^{\rm h} + k^{\rm c} \right)
-\left( k^{\rm h} N^{\rm h} + k^{\rm c} N^{\rm c} \right) N^{\rm h} \right \} \overline{N}^{\rm c}\right.\nonumber\\
&&\qquad\qquad\qquad\qquad\qquad\qquad\qquad\left. - \phi^{\rm h} \left\{ \left( k^{\rm h} + k^{\rm c} \right)
-\left( k^{\rm h} N^{\rm h} + k^{\rm c} N^{\rm c} \right) N^{\rm c} \right \} \overline{N}^{\rm h}
\right] \nonumber\\
&& +\frac{r^2}{1+r^2}
\frac{ 2\left( \phi^{\rm h} -\phi^{\rm c}\right)}{\mathcal{L} }
\frac{k^{\rm h} k^{\rm c} }
{k^{\rm h} \overline{N}^{\rm h} + k^{\rm c} \overline{N}^{\rm c} }
 \left( \phi^{\rm h}k^{\rm h} \overline{N}^{\rm h}
+\phi^{\rm c} k^{\rm c} \overline{N}^{\rm c} \right)  \left( N^{\rm h} - N^{\rm c} \right) \,,
\end{eqnarray}
where $\mathcal{L}$ reads
\begin{eqnarray}
\mathcal{L} =
\left( k^{\rm h} \overline{N}^{\rm h} + k^{\rm c} \overline{N}^{\rm c} \right)
\left( k^{\rm h} \widetilde{N}^{\rm h} + k^{\rm c} \widetilde{N}^{\rm c} \right)
- \left( k^{\rm h} \phi^{\rm h} \overline{N}^{\rm h} +
 k^{\rm c} \phi^{\rm c} \overline{N}^{\rm c} \right)
\left( k^{\rm h} \phi^{\rm h} \widetilde{N}^{\rm h}  +
 k^{\rm c} \phi^{\rm c} \widetilde{N}^{\rm c} \right) \,.
\end{eqnarray}
In the expansion of $\sigma^{\rm q}_d/ \sigma^{\rm cl}_d = \mathcal{S}^0_d + \mathcal{S}^1_d\, \Delta N +
\mathcal{O}(\Delta N^2)$,
the leading terms are given by
\begin{eqnarray} \label{eq:linear1_rconfig}
\mathcal{S}^0_1 = -\frac{r^2}{1+r^2} \frac{2k^{\rm h} k^{\rm c} (1-N^2) }
{\mathcal{L}^{\rm eq}  } \left(\phi^{\rm h} - \phi^{\rm c} \right)^2 \,,\quad
\mathcal{S}^0_2
= -\frac{1}{1+r^2} \frac{2k^{\rm h} k^{\rm c} (1-N^2)  }
{\mathcal{L}^{\rm eq} } \left( \phi^{\rm h} - \phi^{\rm c} \right)^2 \,,
\end{eqnarray}
where
$ \mathcal{L}^{\rm eq} = \left(1-N^2\right) \left[
\left(k^{\rm h} + k^{\rm c}\right)^2 - \left( k^{\rm h} \phi^{\rm h}
+ k^{\rm c} \phi^{\rm c} \right)^2 \right]$.

Setting $\phi^{\rm c} =\phi -\Delta \phi$ and $\phi^{\rm h} =\phi$ with a finite $\phi$,
we obtain
\begin{eqnarray} \label{eq:quadratic1_rconfig}
\mathcal{S}^1_{1} &=& \frac{r^2}{1+r^2}
\frac{2 k^{\rm h} k^{\rm c} \phi \Delta \phi + \mathcal{O}(\Delta \phi^2)}
{ \mathcal{L}^{\rm eq}} + \frac{1}{1+r^2} \frac{2 k^{\rm h} k^{\rm c} \phi \Delta \phi + \mathcal{O}'(\Delta \phi^2)}
{ \mathcal{L}^{\rm eq}} \,\nonumber\\
&\approx&
\frac{2 k^{\rm h} k^{\rm c} \phi \Delta \phi}{ (1-N^2) \left(k^{\rm h} + k^{\rm c}\right)^2
\left(1-\phi^2 \right) } \,, \end{eqnarray}
and
\begin{eqnarray}
\label{eq:quadratic2_rconfig}
\mathcal{S}^1_{2} &=& \frac{1}{1+r^2}
\frac{2 k^{\rm h} k^{\rm c} \phi \Delta \phi + \mathcal{O}(\Delta \phi^2)}
{ \mathcal{L}^{\rm eq}} + \frac{r^2}{1+r^2} \frac{2 k^{\rm h} k^{\rm c} \phi \Delta \phi + \mathcal{O}'(\Delta \phi^2)}
{ \mathcal{L}^{\rm eq}} \,\nonumber\\
&\approx&
\frac{2 k^{\rm h} k^{\rm c} \phi \Delta \phi}{ (1-N^2) \left(k^{\rm h} + k^{\rm c}\right)^2
\left(1-\phi^2 \right) } \,,
\end{eqnarray}
where $\mathcal{O}$ and $\mathcal{O}'$ are higher order terms and
the expansion of $\mathcal{L}^{\rm eq}$ yields
\begin{equation}
\mathcal{L}^{\rm eq} \approx \left(1-N^2\right) \left[ \left(k^{\rm h} + k^{\rm c}\right)^2
\left(1-\phi^2 \right) + 2k^{\rm c} \left( k^{\rm h}
+ k^{\rm c} \right) \phi \Delta \phi \right]\,. \nonumber
\end{equation}
We find that both linear coefficients are negative and $\mathcal{S}^0_1=r^2 \mathcal{S}^0_2$. For $r>1$, the negative contribution from dot
1 (weaker coupling) is stronger. As the second-order coefficients are positive for $\phi \Delta\phi>0$ and stronger than the linear coefficients
for very small $\Delta\phi$, the nonlinear contribution may overcome the linear response to make the quantum conductance positive.

Approaching to $\phi^{\rm h}=\phi^{\rm c} =\pm 1$, the leading order of $\mathcal{L}^{\rm eq}$ becomes
linear in $\Delta \phi$, thus $\mathcal{S}^1_{d}$ remains finite, while
$\mathcal{S}^0_{d}$ goes to zero. Therefore, a strong enhancement of the current is expected.

\section{Fully symmetric case}\label{sec:singular}

We consider the most symmetric case with $\phi^{\rm h}=\phi^{\rm c}=\phi$ in the $r$-symmetric configuration,
where we find simple relations as $W_2=r^2 W_1$, $\overline{W}_2=r^2 \overline{W}_1$, $\Phi=r\phi W_1$,
and $\overline{\Phi}=r\phi \overline{W}_1$. Then, the Liouville matrix becomes
\begin{equation}
\label{eq:Lw1r}
\mathsf{L}= \left(
\begin{array}{ccccc}
-(1+r^2)W_1 & \overline{W}_1 & r^2\overline{W}_1 & r\phi \overline{W}_1 & r\phi  \overline{W}_1 \\
W_1 & -\overline{W}_1 & 0 & \frac{- r\phi \overline{W}_1}{2} & \frac{-r\phi \overline{W}_1}{2} \\
r^2W_1 & 0 & -r^2\overline{W}_1 & \frac{-r\phi \overline{W}_1}{2} & \frac{-r\phi \overline{W}_1}{2} \\
r\phi W_1 &  \frac{-r\phi \overline{W}_1}{2}
& \frac{-r\phi \overline{W}_1}{2} & \frac{-(1+r^2)\overline{W}_1}{2} & 0 \\
r\phi W_1 & \frac{-r\phi \overline{W}_1}{2}
& \frac{-r\phi \overline{W}_1}{2}  & 0 & \frac{-(1+r^2)\overline{W}_1}{2}
 \end{array} \right).
\end{equation}
The eigenvectors and the corresponding eigenvalues of
$\mathsf{L}$ can be obtained from more general results in Sec.~\ref{sec:eigen} or
by directly diagonalizing Eq.~\eqref{eq:Lw1r}. The first three eigenvectors are given as
\begin{align}  \label{eq:eigen_single_asym1}
{\bf v}_1^{\rT} =  \left( \overline{\alpha}, \alpha, \alpha, 0,0 \right) \,,
{\bf v}_2^{\rT} = \left( 0, 0,  0, 1, -1 \right )\,,
{\bf v}_3^{\rT} = \left( 0, 1, -1,\frac{r^2-1}{2r\phi } ,\frac{r^2-1}{2r\phi }  \right ) \,,
\end{align}
where $\alpha={W_1}/({2W_1 + \overline{W}_1})$, $\bar{\alpha} = 1-2 \alpha$, and
the corresponding eigenvalues are $~\lambda_{1}=0, \lambda_2= -\frac{1+r^2}{2}\overline{W}_1$ and $\lambda_3= -\frac{1+r^2}{2} \overline{W}_1$,
respectively. The fourth and the fifth eigenvectors are given as
\begin{align}
\label{eq:eigen_single_asym45}
{\bf v}_{4(5)}^{\rT} = \left(1, -\frac{\lambda_{4(5)} + r^2 (W_1+\overline{W}_1 )-W_1 }
{2\lambda_{4(5) }+(1+r^2) \overline{W}_1 },
 -\frac{\lambda_{4(5)}+ W_1+\overline{W}_1-r^2 W_1 }
{2\lambda_{4(5)} +(1+r^2) \overline{W}_1 },
\frac{r \phi \left(2W_1 +\overline{W}_1\right) }{2\lambda_{4(5)} +(1+r^2) \overline{W}_1 },
\frac{r \phi \left(2W_1 +\overline{W}_1\right) }{2\lambda_{4(5)} +(1+r^2) \overline{W}_1 }
\right)\,
\end{align}
and the corresponding eigenvalues are
\begin{align}
\lambda_{4} = \frac{ -\left(1+r^2 \right) \left( W_1+ \overline{W}_1 \right)+ U}{2}\,,
\quad \lambda_{5} = \frac{ -\left(1+r^2 \right) \left( W_1+ \overline{W}_1 \right)- U}{2}
\end{align}
with $U =\sqrt{ \left[(1+r^2)\left( W_1+\overline{W}_1\right) \right]^2
-4r^2(1-\phi^2)\left(2W_1\overline{W}_1 + \overline{W}_1^2 \right) }$.

One can notice that the maximum interference condition ($|\phi|=1$) yields $\lambda_4=0$, implying
that the steady state is not determined uniquely.  In fact, any state spanned by ${\bf v}_1$
and ${\bf v}_4$ can become a steady state, depending on the initial condition. For $|\phi|< 1$,
all four eigenvalues are negative except $\lambda_1 =0$, so we have a unique steady state
represented ${\bf v}_1$, which is identical to the classical steady state at $\phi=0$.

Defining a matrix $\sf V = ( {\bf v}_1,  {\bf v}_2,  {\bf v}_3,  {\bf v}_4,  {\bf v}_5)$
with its inverse ${\sf V}^{-1}$,
the formal solution $\mathbf{P}(t)$ at time $t$
with an initial vector $\mathbf{P}(0)
= \left(\rho_{00}(0), \rho_{11}(0),\rho_{22}(0),\rho_{12}(0),\rho_{21}(0) \right)^{\rT}$,
reads
\begin{equation}
\label{eq:evolv1}
\mathbf{P}(t) = {\sf V}{\sf V}^{-1} e^{\mathsf{L}\, t}
\,{\sf V}{\sf V}^{-1} \mathbf{P}(0) \,,
\end{equation}
or $\mathbf{P}(t) = {\sf V} \left( 1,  \chi_2 e^{\lambda_2 t},
 \chi_3 e^{\lambda_3 t}, \chi_4 e^{\lambda_4 t}, \chi_5 e^{\lambda_5 t}\right)^{\rT}$,
where we used $\rho_{00}(0) + \rho_{11}(0) + \rho_{22}(0)=1$. All $\chi_i$'s can be calculated
from Eq.~\eqref{eq:evolv1} if the initial condition $\mathbf{P}(0)$ is given.

Let us consider the simple case of $r=1$, where the
eigenvectors and the eigenvalues are given by
${\bf v}_1^{\rT} =  \left( \bar{\alpha}, \alpha ,\alpha , 0,0 \right)$,
${\bf v}_2^{\sf T} = \left( 0,  0, 0, 1 ,-1 \right )$,
${\bf v}_3^{\sf T} = \left( 0, 1,  -1, 0,0 \right )$,
${\bf v}_4^{\sf T} = \left(1, -\frac{1}{2}, -\frac{1}{2},
\frac{U_1+W_1 }{2\phi \overline{W}_1}, \frac{U_1+W_1}{2\phi \overline{W}_1} \right)$,
and ${\bf v}_5^{\sf T} = \left(1, -\frac{1}{2}, -\frac{1}{2},
\frac{W_1-U_1}{2\phi \overline{W}_1}, \frac{W_1-U_1}{2\phi \overline{W}_1} \right)$,
with $\lambda_{1}=0$,
$\lambda_2= -\overline{W}_1$,
$\lambda_3= -\overline{W}_1$,
$\lambda_4 = -(W_1 + \overline{W}_1) + U_1$
and $\lambda_5 = -(W_1 + \overline{W}_1) - U_1$ ,
where
$U_1=\sqrt{( W_1+\overline{W}_1)^2
-(1-\phi^2) 2W_1\overline{W}_1 + \overline{W}_1^2 )}$.
Then the inverse matrix $\mathsf{V}^{-1}$ is obtained as
\begin{equation}
\mathsf{V}^{-1}= \left(
\begin{array}{ccccc}
1 &1 &1 &0 &0 \\
0&0&0& \frac{1}{2} & -\frac{1}{2} \\
0&\frac{1}{2}&-\frac{1}{2}&0&0 \\
\frac{\alpha (U_1 - W_1)}{U_1} & -\frac{\bar\alpha (U_1 - W_1)}{2U_1} &
 -\frac{\bar\alpha (U_1 - W_1)}{2U_1} & \frac{\phi \overline{W}_1}{2U_1} &  \frac{\phi \overline{W}_1}{2U_1}\\
\frac{\alpha (U_1 + W_1)}{U_1} & -\frac{\bar\alpha (U_1 + W_1)}{2U_1} &
 -\frac{\bar\alpha (U_1 + W_1)}{2U_1} & -\frac{\phi \overline{W}_1}{2U_1} &  -\frac{\phi \overline{W}_1}{2U_1}
\end{array} \right) \,.
\end{equation}
If the initial condition is given by $\rho_{11}(0)=\rho_{22}(0)$ and $\rho_{12}(0)=\rho_{21}(0)$,
Eq.~\eqref{eq:evolv1} is simplified since $\chi_2=\chi_3 =0$ and thus
$\rho_{11}(t)= \rho_{22}(t)$ and $\rho_{12}(t) = \rho_{21}(t)$.
It is straightforward to obtain the dynamic equation
\begin{equation}
\label{eq:evolv2}
\left( \begin{array}{c}
\rho_{11}(t) \\ \rho_{12}(t) \end{array} \right) = \alpha
\mathbf{A}(\phi, t) + \mathsf{M}(\phi, t)
\left( \begin{array}{c}
\rho_{11}(0) \\ \rho_{12}(0) \end{array} \right)\,,
\end{equation}
where the vector $\mathbf{A}$ is given by
\begin{equation}
\mathbf{A}(\phi, t) = \frac{1}{2} \left( \begin{array}{c}
2-R_{+}(\phi,t) + \frac{W_1}{U_1 }  R_{-}(\phi,t) \\
 \frac{\phi(2W_1 + \overline{W}_1) }{ U_1}
R_{-}(\phi,t) \end{array} \right)\,,
\end{equation}
and the Matrix $\mathsf{M}$,
\begin{equation}
\mathsf{M}(\phi,t) = \frac{1}{2}\left(
\begin{array}{cc}
R_{+}(\phi,t) - \frac{W_1}{U_1 }R_{-}(\phi,t) &
-\frac{\phi \overline{W}_1} {U_1 } R_{-}(\phi,t)\\
-\frac{\phi(2W_1 + \overline{W}_1) }{ U} R_{-}(\phi,t)
&R_{+}(\phi,t) + \frac{W_1 }{U_1 }R_{-}(\phi,t)
\end{array} \right) \,.
\end{equation}
Here, $R_{\pm}(\phi,t)= e^{ \lambda_4 t}
\pm e^{ \lambda_5 t}$. The data in Fig.~\ref{fig:3}  of the main text are
calculated from Eq.~\eqref{eq:evolv2} with the initial state of $\rho_{11}(0)=\rho_{12}(0)=0$.

At the singular point ($\phi=1$), the multiple steady states emerge, depending on the initial condition.
As the two eigenvalues ($\lambda_1=\lambda_4$) become zero, there should be a conservation law associated
with $\lambda_4$, in addition to the probability conservation responsible for $\lambda_1$. From
the structure of the Liouville matrix in Eq.~\eqref{eq:Lw1r}, one can easily find the conservation law
of $r^2\dot{\rho}_{11} + \dot{\rho}_{22} - r\dot{\rho}_{12} - r\dot{\rho}_{21}=0$ for $\phi=1$. This implies that
the quantity $r^2\rho_{11}(t) + \rho_{22}(t) - r\rho_{12}(t) - r\rho_{21}(t)=I_0$ does not change in time.
Using the relation of Eq.~\eqref{eq:sol_ss1} of the main text and this conservation law, we obtain
the multiple fixed points as
\begin{align}
&\rho_{11}(\infty) = \alpha -
[r \bar{\alpha} -\frac{1-r^2}{r}\alpha]\, \rho_{12}(\infty) \,,\quad
\rho_{22}(\infty) = \alpha -[ \frac{\bar{\alpha}}{r} + \frac{1-r^2}{r} \alpha ]\, \rho_{12}(\infty) \,,\nonumber\\
\label{eq:phi+1}
&\rho_{12}(\infty) =\rho_{21}(\infty)=  \frac{r }{1+r^2}
\frac{1}{1-\alpha} \left(  \alpha - \frac{I_0}{1 + r^2} \right)~,
\end{align}
where $I_0=r^2 \rho_{11}(0)+ \rho_{22}(0) - r\rho_{12}(0) - r\rho_{21}(0)$.
Similarly, we get the extra conservation of $r^2\rho_{11}(t) + \rho_{22}(t) + r\rho_{12}(t) + r\rho_{21}(t)=I_0^\prime$ for $\phi=-1$
and the corresponding multiple fixed points are given as
\begin{align}
&\rho_{11}(\infty) = \alpha +
[r \bar{\alpha} -\frac{1-r^2}{r}\alpha]\, \rho_{12}(\infty) \,,\quad
\rho_{22}(\infty) = \alpha +[ \frac{\bar{\alpha}}{r} + \frac{1-r^2}{r} \alpha ]\, \rho_{12}(\infty) \,,\nonumber\\
\label{eq:phi-1}
&\rho_{12}(\infty) =\rho_{21}(\infty)=  -\frac{r }{1+r^2}
\frac{1}{1-\alpha} \left(  \alpha - \frac{I_0^\prime}{1 + r^2} \right)~,
\end{align}
where $I_0^\prime=r^2 \rho_{11}(0)+ \rho_{22}(0) + r\rho_{12}(0) + r\rho_{21}(0)$.

We can also calculate the steady-state currents.
From Eqs.~\eqref{eq:Jcl1}-\eqref{eq:Psi2}, we obtain
\begin{eqnarray}
\Psi_1 &=& \frac{\overline{\Phi}}{|\mathsf{L}_0|}
(1+r^2)\left( w^{\rm h}_{1+} \overline{W}_1 - w^{\rm h}_{1 -} W_1 \right)
=\phi\frac{1+r^2}{r} J_1^{\rm cl}\,, \nonumber\\
 \Psi_2 &=& \frac{\overline{\Phi}}{|\mathsf{L}_0|} \left(1+\frac{1}{r^2}\right)
\left( w^{\rm h}_{2+} \overline{W}_2 - w^{\rm h}_{2-} W_2 \right)
=\phi\frac{1+r^2}{r} J_2^{\rm cl}\,,\label{eq:Psi}
\end{eqnarray}
where
\begin{align}
J_1^{\rm cl}=\frac{2\pi |g_1^{\rm h}|^2 |g_1^{\rm c}|^2 \Delta N}
{|g_1^{\rm h}|^2 \widetilde{N}^{\rm h}+ |g_1^{\rm c}|^2 \widetilde{N}^{\rm c}}\,, \quad\quad J_2^{\rm cl}=r^2 J_1^{\rm cl} \,,
\end{align}
with $\widetilde{N}^{a}=1+N^{a}$.
For $|\phi|<1$, the classical solution becomes the unique steady state ($\rho_{12}(\infty)=0$), thus
the steady-state current is purely classical. However, at $|\phi|=1$, we have a nonzero $\rho_{12}(\infty)$
in Eqs.~\eqref{eq:phi+1} and \eqref{eq:phi-1} and
the steady-state current
contains the quantum part as $J_d(\infty)=J_d^{\rm cl} +\Psi_d \rho(\infty)$, thus, for $\phi=\pm 1$,
\begin{align}
J_{d}(\infty) = J_{d}^{\rm cl} \left( 1\pm \frac{1+r^2}{r}\, \rho_{12}(\infty) \right)\,.
\end{align}
Note that these currents are the same for $\phi=\pm 1$ with the initial conditions of $I_0=I_0^\prime$.
In equilibrium, the quantum current vanishes as well as the classical current even if $\rho_{12}(\infty)\neq 0$,
because the quantum speed $\Psi_d$ vanishes at $\Delta N=0$ as in Eq.~\eqref{eq:Psi}.